\documentclass[twocolumn]{aastex631}
\usepackage{natbib,graphics,graphicx,multirow,amsmath,cancel,color,xcolor,hyperref}
\usepackage{soul}
\usepackage[normalem]{ulem}
\usepackage{float}
\usepackage{comment}
\usepackage{tabularx}
\usepackage{multirow}
\usepackage{array}
\usepackage{enumitem}
\renewcommand{\arraystretch}{1.2}
\usepackage{color, colortbl}
\usepackage{bm}

\begin{document}
\title{Characterizing Nanoflare Energy and Frequency through Field Line Analysis}

\correspondingauthor{Shanwlee Sow Mondal}
\email{shanwlee.sowmondal@nasa.gov\\
shanwlee.sowmondal@gmail.com}

\author[0000-0003-4225-8520]{Shanwlee Sow Mondal}
\affil{Department of Physics, The Catholic University of America, 620 Michigan Ave., N.E. Washington, DC 20064, USA}
\affil{Heliophysics Science Division, NASA Goddard Space Flight Center, 8800 Greenbelt Rd., Greenbelt, MD 20771, USA}

\author[0000-0002-1198-5138]{L. K. S. Daldorff}
\affil{Department of Physics, The Catholic University of America, 620 Michigan Ave., N.E. Washington, DC 20064, USA}
\affil{Heliophysics Science Division, NASA Goddard Space Flight Center, 8800 Greenbelt Rd., Greenbelt, MD 20771, USA}

\author[0000-0003-2255-0305]{James A. Klimchuk}
\affil{Heliophysics Science Division, NASA Goddard Space Flight Center, 8800 Greenbelt Rd., Greenbelt, MD 20771, USA}

\author[0000-0003-4023-9887]{C. D. Johnston}
\affil{Physics and Astronomy Department, George Mason University, 4400 University Dr, Fairfax, VA 22030, USA}
\affil{Heliophysics Science Division, NASA Goddard Space Flight Center, 8800 Greenbelt Rd., Greenbelt, MD 20771, USA}

\begin{abstract}
We present a detailed analysis of a 3D MHD simulation of a subset of the magnetic flux in an active region. The simulation models the generation of nanoflares and response of the plasma to imposed photospheric motions. Our study focuses on characterizing the energy distribution and occurrence frequency of the nanoflares in the simulation that self-consistently heat the corona. This field line based analysis reveals that the nanoflare energy distribution (energy per unit cross sectional area) follows a log-normal profile, where low energy nanoflares are significantly more prevalent than those with high energy. When compared with the plasma cooling time, different energy nanoflares tend to repeat with different frequencies. 
Low energy nanoflares repeat at high frequencies, while high energy nanoflares repeat at low frequencies. 
However, the thermal evolution of plasma along individual field lines is governed predominantly by the high energy nanoflares. 
These findings provide critical insights into the role of small-scale magnetic reconnection events in heating the solar corona.
\end{abstract}


\section{Introduction}\label{sec:introduction}
Coronal heating remains a major unsolved problem in solar physics. It is widely believed that the energy originates from complex photospheric motions that twist and tangle coronal field lines. To prevent infinite magnetic stress buildup, the field undergoes reconnection \citep{Klimchuk_2006, Klimchuk_2015}, releasing energy through small impulsive events called `nanoflares' \citep{Parker_1983, Parker_1988, Klimchuk_2006, Klimchuk_2015}. Because nanoflares are small and overlap along the optically thin line-of-sight, they are unresolvable as individual events by current instrumentation. 

An important property to understand the underlying heating mechanism is the occurrence frequency of nanoflares on a given magnetic strand.  Magnetic strands are the building blocks of the low-$\beta$ corona. They refer to magnetic substructures with approximately isothermal cross-sections \citep{Klimchuk_2015}. In the case of low-frequency nanoflares, the delay between consecutive nanoflares is long compared to the characteristic cooling time. Hence, the strand fully cools and drains before it is re-energized. In contrast, high-frequency nanoflares repeat with delays shorter than the cooling time and produce heating conditions that are effectively `steady'. Nanoflares with intermediate occurrence frequencies repeat with delays of the order of the plasma cooling time. 
 
Because nanoflares cannot be observed individually, observational studies have examined the collective emission from multiple unresolved events. Emission measure slopes \citep{Tripathi_2011ApJ...740..111T,Warren_2011ApJ...734...90W,Warren_2012ApJ...759..141W,Winebarger_2011ApJ...740....2W,Schmelz_2012ApJ...756..126S,DelZanna_2015A&A...573A.104D} and time lag analysis \citep{Viall_2012ApJ...753...35V, Viall_2017ApJ...842..108V} have been especially useful for inferring nanoflare frequencies. Results suggest a broad distribution of frequencies in active regions, with a characteristic delay comparable to the loop cooling time ($\sim 1000$s) \citep{Cargill_2015RSPTA.37340260C,Barnes_2021ApJ...919..132B}.

Several studies have used hydrodynamic simulations, i.e. `loop models', to infer nanoflare frequencies through the comparison of synthetic and observed emission measure distributions \citep[e.g.,][]{Mulu-Moore_2011ApJ...742L...6M,Bradshaw_2012ApJ...758...53B,Reep_2013ApJ...764..193R}. However, the delay between successive nanoflares has always been an ad hoc input to these models. For example, \citet{Reep_2013ApJ...764..193R} assumed a constant delay, while later studies assumed a relationship between the nanoflare delay and energy, with both obeying power-law distributions \citep{Cargill_2014ApJ...784...49C,Cargill_2015RSPTA.37340260C,Barnes_2016ApJ...833..217B}.

Comparisons between models and observations of active region cores have led to seemingly contradictory results. \citet{Barnes_2019ApJ...880...56B,Barnes_2021ApJ...919..132B} and \citet{Warren_2020ApJ...896...51W} concluded that  heating at the core is high frequency, while \cite{Mondal_2025ApJ...980...75M} concluded that it is low frequency. 
The differences here could be due to real differences in the active regions or it could be due to the differing assumptions made about the shape and range of the nanoflare energy distribution. 

The hydro-models are effective for studying field-aligned plasma but they cannot self-consistently simulate the magnetic field evolution and plasma response. 
In fact, self-consistent simulations of coronal heating are extremely challenging because of the enormous range of spatial and temporal scales that are involved, and the coupling between them. Coronal loops, which appear as bright, distinct features against their surroundings, consist of numerous magnetic strands bundled together. 
The strands are separated by thin current sheets that are formed through random motions in the photosphere displacing magnetic field lines. Observations indicate that a single active region contains more than $10^{5}$ magnetic strands and hence a similar number of current sheets \citep{Klimchuk_2015}. 
The most advanced numerical grids currently available can only resolve a small fraction of this large number of thin current sheets that are found in an active region. Existing MHD models of entire solar active regions therefore include only large volume currents which are significantly different from the actual thin current sheets that are present in active regions \citep{Gudiksen_2002astro.ph..3167G,Bourdin_2013A&A...555A.123B,Warnecke_2019A&A...624L..12W,Martinez_2011ApJ...743...23M,Rempel_2017ApJ...834...10R}. 

This has led to the development of a new class of models called \textit{multi-strand} models \citep{Johnston_2025arXiv250812952J}. These models involve fewer magnetic strands that result in fewer current sheets to facilitate detailed studies of coronal heating over small subsets of active regions.

\cite{Bingert_2013A&A...550A..30B} presented a statistical approach to analyze heating events in a 3D MHD model of an active region. The analysis reveals a power-law distribution for heating event energies where the heating is due to ohmic dissipation. \cite{Kanella_2017A&A...603A..83K,Kanella_2018A&A...617A..50K,Kanella_2019A&A...621A..95K} also measured the distribution of volume-integrated energies of reconnection events. They proposed a method of identifying 3D heating events and found that the energy released per event followed a power-law with a slope of about $-1.5$. The authors used ohmic heating as a proxy to identify reconnection events. Existing multi-strand studies    \citep{Galsgaard_1996JGR...10113445G,Rappazzo_2013ApJ...771...76R,Reid_2020A&A...633A.158R,Howson_2022A&A...661A.144H,Breu_2022A&A...658A..45B} have provided extensive insights into various aspects of coronal heating, such as the effect of driving time scales and energy release through the kink instability and avalanches. However, detailed studies on the heating frequency of nanoflares remain limited.

Recently, \cite{Knizhnik_2020SoPh..295...21K} used MHD simulations to study the distribution of delays between successive nanoflares. They found that nanoflare delays follow a power-law distribution with slope $\sim -1$ in the range $10^{2}-10^{4}$ s. The inferred slope is much shallower than previously reported from observations and determined from loop model comparisons with observations.

Our goal is to study the energy distributions and occurrence frequencies of nanoflares in a 3D MHD simulation of a small subset of an active region. This study differs from conventional nanoflare simulations by employing a field-line-based approach to investigate heating driven by small-scale reconnection events along individual field lines, rather than attributing heating to large-scale volumetric currents or flows.
Nanoflare energies in this study are expressed as heating integrated along the field line and quantified per unit area. In contrast, previous studies typically used proxies to identify reconnection events and calculated the energy released over the entire volume of the event. Energy per unit area is more appropriate for understanding how the plasma responds to the energy input and produces the spectrum of emitted radiation.

The rest of the paper is organized as follows. Section \ref{sec:simulation details} details the simulation setup. Section \ref{sec:Methodology} outlines the methods used to quantify nanoflare energies associated with each field line. Section \ref{sec:Results} compares the results from these different methods. Lastly, Section \ref{sec:Summary} discusses and summarizes the main findings of this study.

  \begin{figure*}
  \centering
      \includegraphics[width=0.95\textwidth,angle=0]{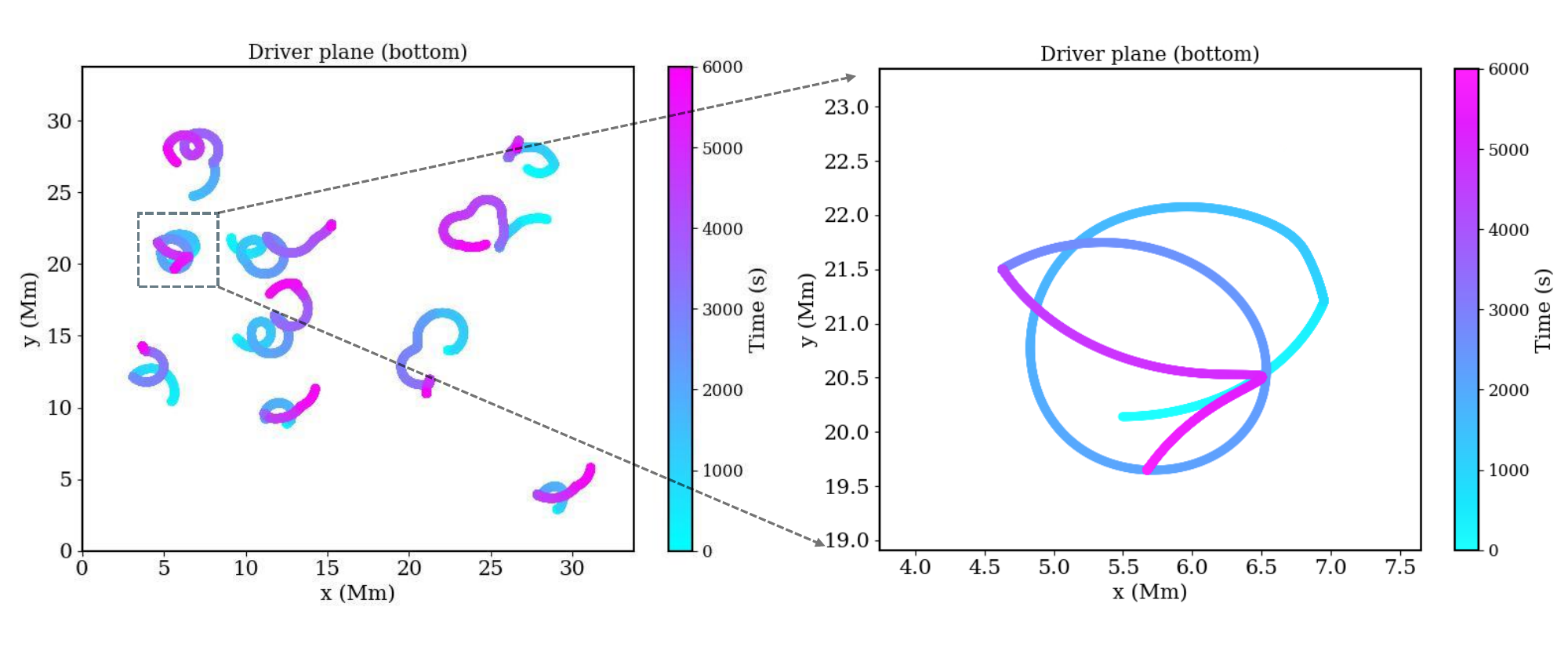}
 \caption{Trajectories of 10 randomly chosen footpoints in the bottom photospheric driving plane. The color bar shows the evolution of time, with Time=0 denoting the time when we start our analysis ($\sim$ 350 minutes after the start of the simulation), which corresponds to a time when the system has already reached a statistical steady state. The zoomed-in trajectory of the footpoint indicated by the dashed box on the left is displayed in the right plot. The trajectories clearly show the rotational and translational aspects of our photospheric driving.}
  \label{trajectories}
  \end{figure*}

\section{Simulation details} \label{sec:simulation details}

A detailed description of the simulation is presented in \cite{Johnston_2025arXiv250812952J}, so we just provide a brief description below.

\subsection{Numerical Setup}

The following MHD equations are solved using the Lagrangian Remap (LaRe) code \citep{Arber_2001JCoPh.171..151A}:

\begin{eqnarray}
    \frac{\partial \rho}{\partial t}  
    + \nabla \cdot (\rho \mathbf{v}) = 0
    \\
    \rho \frac{D \mathbf{v}}{D t} 
    = - \nabla P + \rho \mathbf{g} + \mathbf{j} \times \mathbf{B} + \mathbf{F}_{\textrm{shock}}
    \\
    \frac{\partial \mathbf{B}}{\partial t}
    = \nabla \times (\mathbf{v} \times \mathbf{B})
    \\
    \rho \frac{D \epsilon}{D t} 
    = - P \nabla \cdot \mathbf{v} 
    + Q_{\textrm{shock}} 
    - \nabla \cdot \mathbf{q} 
    - n^{2} \Lambda(T)
    \\
    P = 2 k_{B} n T
\end{eqnarray}

\noindent where $\frac{D}{Dt} = \frac{\partial}{\partial t}+ (\bf{v}\cdot\nabla)$,  $\rho$ is the mass density, $n$ the electron number density where we have assumed a fully ionized hydrogen plasma, $\bf{v}$ the plasma velocity, $T$ the temperature, $k_{B}$ the Boltzmann constant, $\bf{B}$ the magnetic field, $P$ the gas pressure, $\epsilon = P/\rho (\gamma-1)$ the specific internal energy (where $\gamma = 5/3$ is the ratio of specific heats), ${\bf j=(\nabla \times B})/\mu_0$ the current density, $\bf{g}$ the gravity, and $\bf{q}$ the heat flux vector. ${\bf F}_{\textrm{shock}}$ is the force per unit volume from the shock viscosity and $Q_{\textrm{shock}}$ is the associated shock heating. No other form of background heating is imposed. 
LaRe3D captures the transfer of reconnection energy as it is expected to occur on the Sun – from magnetic, to kinetic, to thermal at shocks.
$\Lambda(T)$ is the radiative loss function for an optically thin plasma, which we approximate using the function defined in \cite{Klimchuk_2008ApJ...682.1351K}. The radiative loss function is modified at low temperatures to produce an approximately isothermal chromosphere at $2\times 10^{4}$ K \citep{Klimchuk_1987ApJ...320..409K}. 
The thermodynamic coupling between the corona, transition region (TR), and chromosphere is captured using the Transition Region Adaptive Conduction (TRAC) method \citep{Johnston_2019ApJ...873L..22J,Johnston2020,Johnston_2021A&A...654A...2J}, which allows us to accurately treat the plasma response to impulsive heating events using fewer grid points than would otherwise be required to fully resolve the TR.

\textit{Domain size -- }The model represents a small subset of magnetic flux within an active region. The simulation domain spans $33.75 \times 33.75$ Mm$^2$ in the horizontal directions and $130$ Mm vertically.  The effective `loop' length measured between the two chromospheres is $100$ Mm. The photospheric boundaries are situated 6 Mm below the top of the chromosphere on both sides. This is 9 Mm above the boundaries of the numerical box, for reasons discussed below. The computational domain is uniformly gridded with $256 \times 256 \times 512$ points along the $x$, $y$ and $z$ directions, respectively. See \cite{Johnston_2025arXiv250812952J} for more details. 

\textit{Initial and boundary conditions -- } 
The simulation begins with a Parker atmosphere \citep{Parker_1972} where the initial state is gravitationally stratified with realistic temperature and density profiles that correspond to a photosphere - chromosphere - transition region - corona system. 
A uniform magnetic field of 100 G, oriented along the vertical axis, is imposed throughout the domain. The top and bottom of the box correspond to the positive and negative polarity regions of the photosphere. Gravity along the field line is analogous to that of a semicircular loop - with the coronal mid-plane serving as the loop apex where gravity vanishes - and is directed downward toward the photospheric boundaries on each side. This is intended to represent the curved field of an active region arcade. The system is symmetric about the loop apex, with plasma $\beta = 0.001$ in the corona. The boundary conditions are periodic in the $x$ and $y$ directions, and a prescribed velocity profile is applied at the photospheric boundaries. 

\begin{figure*}
  \centering
      \includegraphics[width=0.95\textwidth,angle=0]{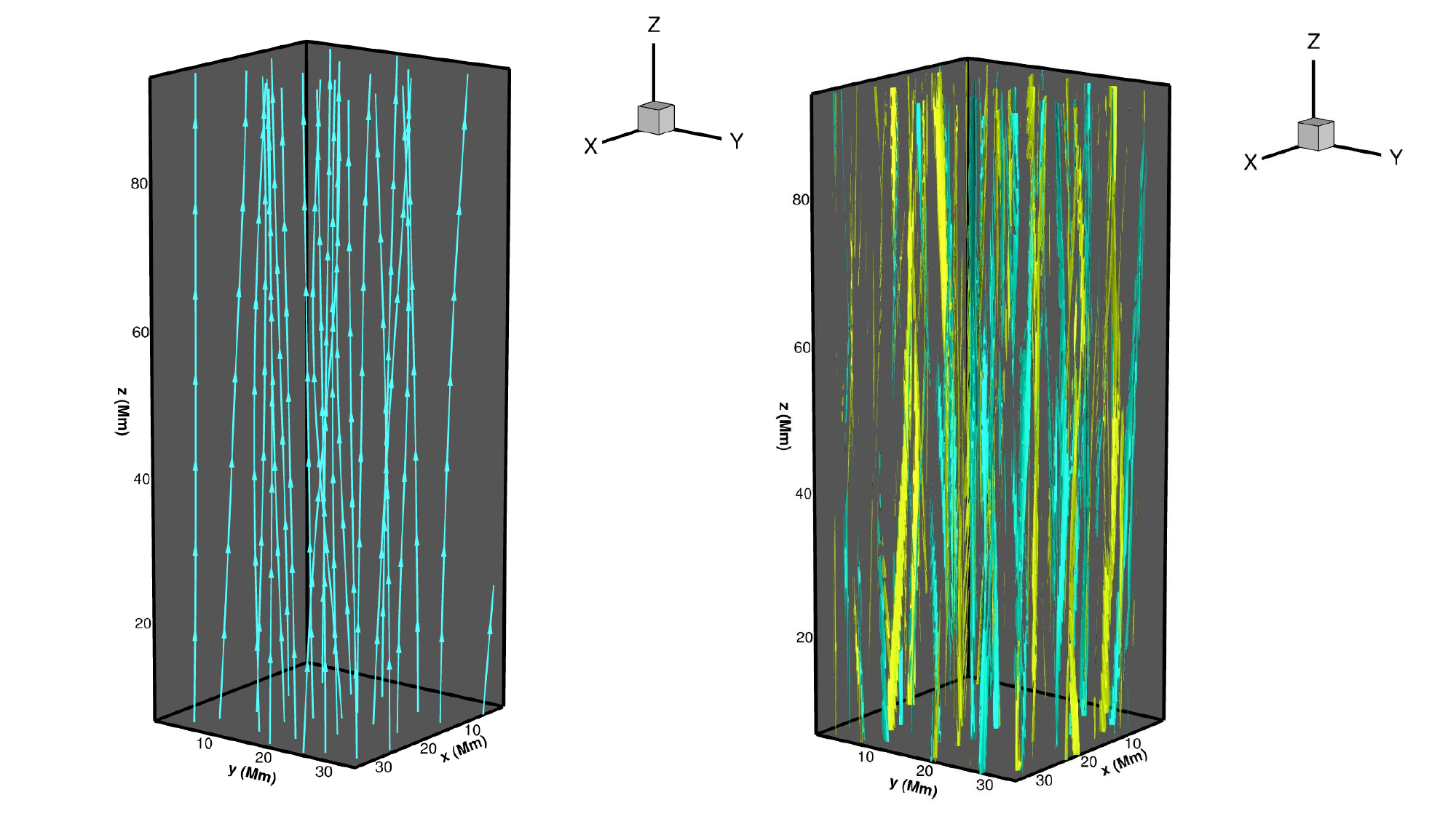}
 \caption{(left) Random field lines and (right) $\pm 2\times10^{-3}$ Amp/m$^{2}$ iso-surfaces of vertical current density in the multi-strand simulation. The elongated current sheets are formed at the boundaries of misaligned quasi-homogeneous flux strands produced by driving at both ends.}
  \label{fieldlines}
  \end{figure*}

\subsection{Photospheric driving }
We drive the magnetic field at a depth of 6~Mm below the top of the chromospheres, corresponding to two photospheric boundaries. This is above the $z$ boundaries of the box to avoid unwanted boundary condition effects (see e.g., Daldorff et. al, in preparation).
The velocity profile is meant to represent observed photospheric motions. The driving pattern consists of two $6\times6$ arrays of rotational cells. Each array fades in and out over a 20 minute cycle. They are half a cycle out of phase, and each array is translated randomly after its cycle is completed. The resulting pattern resembles a random walk with curved steps. Figure \ref{trajectories} shows the streamlines of 10 randomly chosen footpoints at the bottom driver plane. The flows are anti-parallel at the top plane. Each vortex cell has a radial profile of velocity with a maximum value of $7.5$ km/s. This driving speed magnitude is chosen so that all footpoint motions are sub-Alfv\'enic. 
The vertical magnetic field component remains uniform in the photospheric driving planes due to the incompressible nature of the flow. On the other hand, the horizontal component is free to evolve in response to the imposed driving.

\subsection{Generation of current sheets and energy release}
Photospheric flows twist and tangle the coronal magnetic field, leading to the buildup of magnetic energy. Thin current sheets form at the boundaries of quasi-homogeneous magnetic flux strands. Figure \ref{fieldlines} illustrates the complexity of the field and shows iso-surfaces of vertical current density between the photospheric driving planes. To prevent infinite stress accumulation, the magnetic strands eventually reconnect, releasing the stored magnetic energy in small but impulsive events. The system eventually reaches a statistical steady state when the injected Poynting flux balances the energy released during the impulsive events. While individual events are irregular and unpredictable, the overall energy balance remains steady, keeping the system's macroscopic properties (e.g., average temperature and magnetic field strength) unchanged over time. Although not shown here, the simulation generates coronal temperatures of several million Kelvin and produces emission patterns similar to observations as described in detail in \cite{Johnston_2025arXiv250812952J}. 

After the system has reached its statistical steady state, we apply the following methods to analyze the heating characteristics of the nanoflares in the simulation.

\section{Methodology} \label{sec:Methodology}
To quantify the frequency of nanoflares occurring on each magnetic field line, we tracked the footpoints of multiple field lines over time. 
This is done using the following method.

\subsection{Field line tracking \& measuring $\delta$}
\begin{figure}
  \centering
     \includegraphics[width=0.49\textwidth,angle=0]{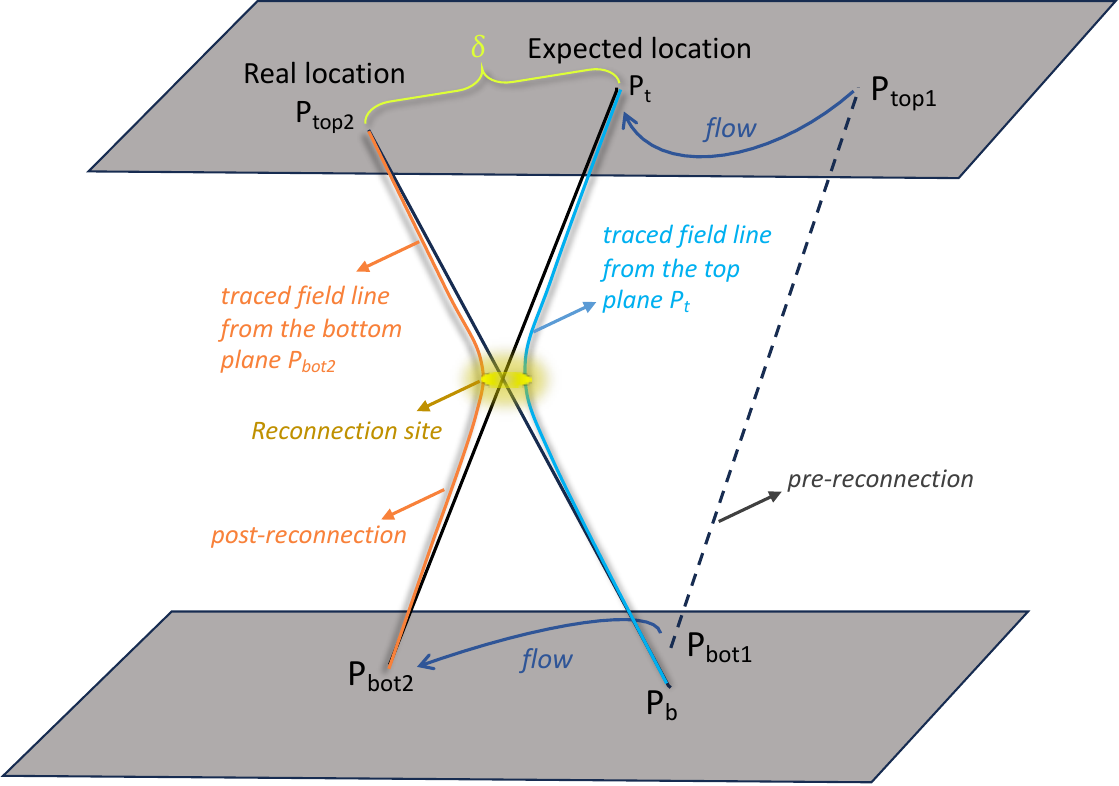}
      \caption{Sketch showing the method of determining $\delta$ -- a quantity to identify if a field line has undergone reconnection.}
      \label{Cartoon}
\end{figure}

Consider a field line with its footpoint at the bottom driving plane, denoted as $P_{bot1}$, at time $t_{1}$ (see Figure \ref{Cartoon}) which maps to $ P_{top1}$ at the top driving plane. This mapping is done with our in-house developed \textit{field line tracing} algorithm which uses a fourth/fifth-order Runge-Kutta-Fehlberg scheme for field line tracing. Given the known driving patterns on both planes, we estimate the advected positions of $P_{bot1}$ and $P_{top1}$ at time $t_{2}$ which are given by $P_{bot2}$ and $P_{t}$, respectively. When $P_{bot2}$ is mapped to the top plane, it lands at location $P_{top2}$, which will be different from $P_{t}$ if reconnection has occurred. We then measure the distance between the mapped and advected locations at the top plane as a function of time --

\begin{equation}
  \delta(t_{2}) = \big|\underbrace{P_{top2}}_{\textit{real location}} - \underbrace{[P_{top1}+\int_{t_1}^{t_2}\bm{v}_{top}(t) dt]}_{\textit{expected location}}   \big|
\end{equation}

\noindent where $\bm{v}_{top}(t)$ is the driving velocity in the top plane at time $t$, $\Delta t =t_{2}-t_{1}=1$ s, and $dt( << \Delta t$) is the integration step size.

In an ideal case with no reconnection, we expect $\delta = 0$, which means that the evolution of the field line is purely ideal and is simply advected by the flow. Conversely, if reconnection occurs, the change in magnetic connectivity, i.e., the mapping between the driving planes of the field line, gives rise to $\delta > 0$. Figure \ref{Cartoon} provides a pictorial representation of how $\delta$ is estimated for a footpoint at any given time. A similar approach was also used by \citep{Knizhnik_2020SoPh..295...21K}.

\subsection{Reconnection-driven $\delta$} 
In addition to reconnection, other potential factors for non-zero $\delta$s can be numerical diffusion and small errors in field line tracing. Hence, to isolate reconnection driven non-zero $\delta$s from other sources, we have used a threshold condition such that any $\delta$, satisfying $\delta \geq \delta_{cut}$, is considered as solely reconnection driven, whereas $\delta < \delta_{cut}$ contains the degeneracy from all the above mentioned possible factors. In this study, we have considered $\delta_{cut} =1$, $2$ and $3$ grid cells.

\subsection{Coronal reconnections} 
\label{sec:coronal_recn}
Chromospheric heating events are minor contributors to observed coronal emission \citep{Klimchuk_2012JGRA..11712102K,Klimchuk_2014ApJ...791...60K,Bradshaw_2015ApJ...811..129B,SowMondal_2022ApJ...937...71S}. Transition region heating events are not considered, as we expect them to make a relatively small contribution to total heating, as borne out in our simulation. Therefore, this study includes reconnections that occur only at coronal heights. To identify coronal reconnections, we need to pinpoint the exact location of reconnection along a field line, which is beyond the scope of our present study. Instead, we assume that the location along a field line where the volumetric viscous heating rate ($H_{visc}$) reaches its maximum value corresponds to the reconnection site. This assumption is based on the expectation that strong flows, and consequently strong shock heating -- captured by the volumetric viscous heating term -- are present near reconnection sites. Cases in which the maximum $H_{visc}$ is located at coronal heights (at least 3 Mm above the initial chromospheric height) are considered relevant for coronal heating. In contrast, cases where the maximum $H_{visc}$ is outside the defined coronal range contribute less to coronal emission and are therefore excluded from this study.

\begin{figure}
 \centering
 \includegraphics[width=0.5\textwidth, angle=0]{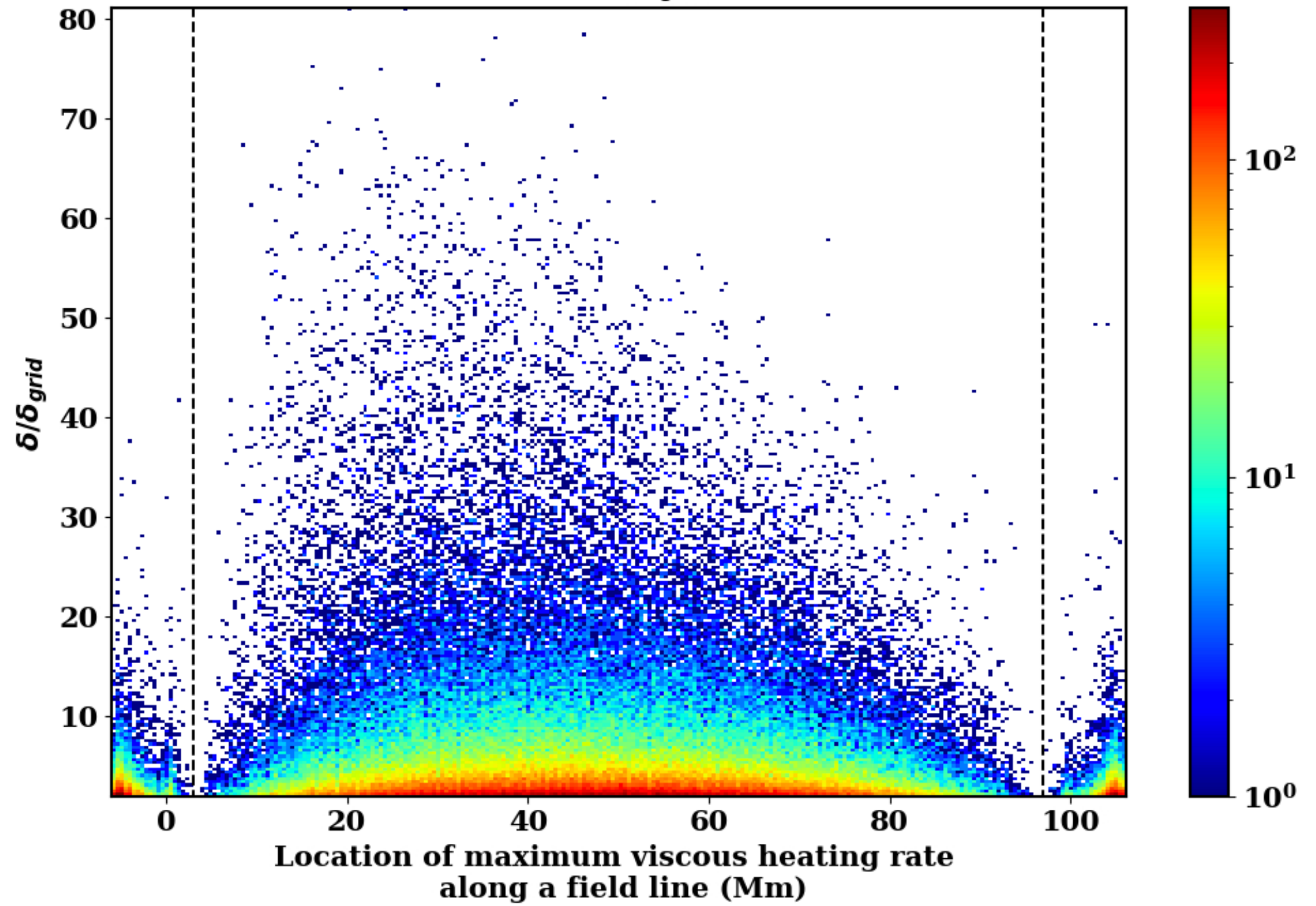}
 \caption{2D histogram of $\delta$ jumps versus the location of maximum viscous heating rate along a field line obtained from 3136 footpoints tracked over 6000 s.}
  \label{delta_vs_ht}
  \end{figure}
 
\begin{figure*}
 \centering
 \includegraphics[width=0.95\textwidth,angle=0]{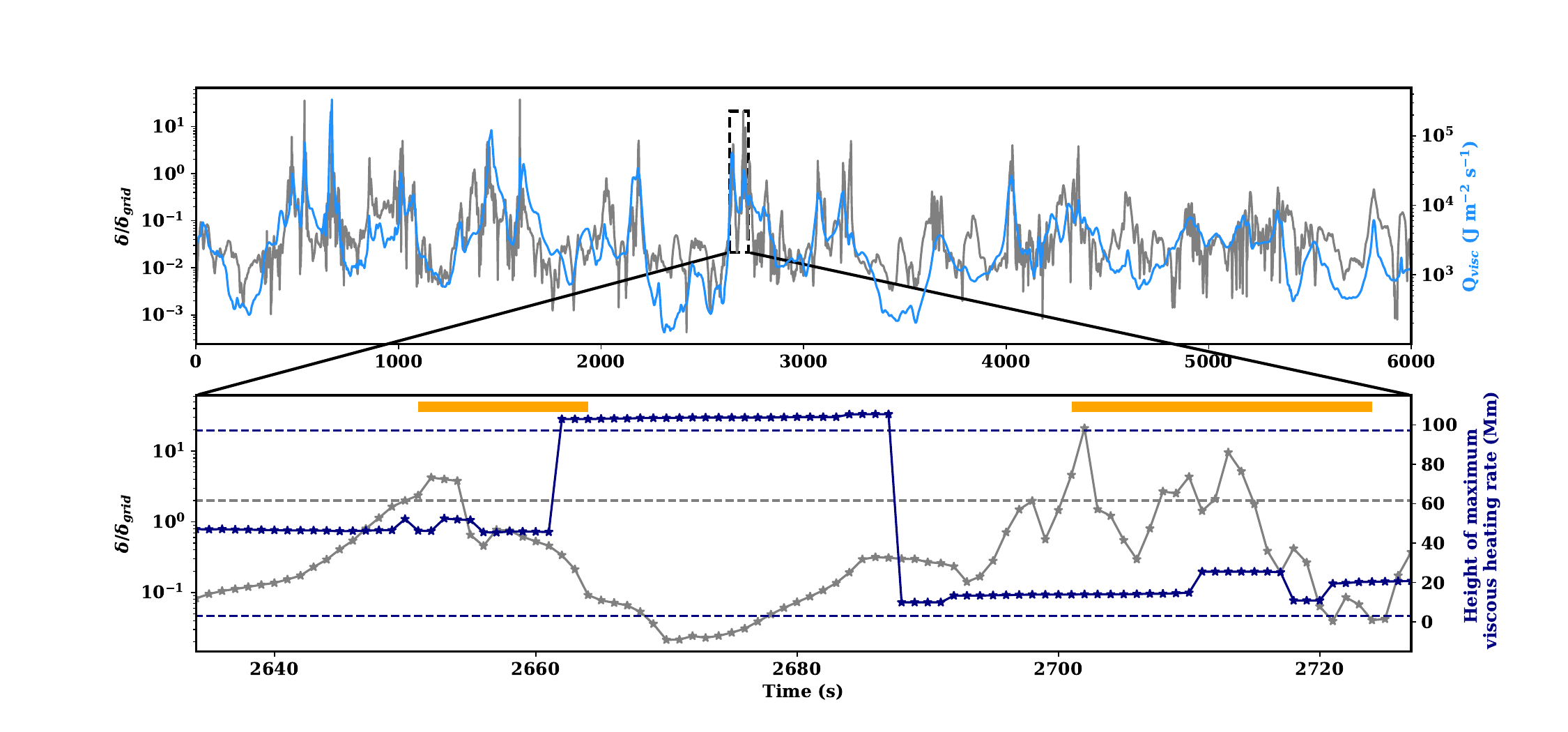}
 \caption{Top panel: Evolution of reconnection jump ($\delta$) in grey and coronal integrated viscous heating rate in blue for a randomly chosen footpoint. The $\delta$s are normalised with the grid resolution ($\delta_{grid}$). Bottom panel: Normalized $\delta$s (in grey) and the corresponding heights of maximum viscous heating rate (in blue) for the chosen field line in the chosen duration marked by the dashed box in top panel. The grey dashed line marks $\delta_{cut} = 2$ grid cells, while the blue horizontal lines indicate the coronal boundaries. The two orange bars show the durations of two nanoflares as defined in the main text using $t_{window} = 10$ s.}
\label{delta_vs_time}
\end{figure*}

Over a duration of 6000 s, we have tracked 3136 field lines with initial footpoints distributed randomly over the bottom driver plane. We have calculated $\delta$ for each footpoint at 1 s intervals. Figure \ref{delta_vs_ht} presents a 2D histogram of $\delta$ ($ > 2$ grid cells) and the location of maximum $H_{visc}$ along a field line as measured from the top of the left chromosphere. The left and right ends of the x-axis represent the two driving planes, while the two vertical lines -- at 3 Mm and 97 Mm -- indicate the coronal boundaries. This study considers all data points located between these two vertical lines. The histogram indicates that the number of reconnections increases with height in the corona. There are also significant reconnections between the driver plane and transition region, but there are relatively few reconnections in the transition region itself. This justifies our neglect of these events in our analysis. These trends are especially strong for the larger $\delta$ jumps.

Figure \ref{delta_vs_ht} shows a slight asymmetry, with maximum viscous heating more frequently occurring preferentially closer to the lower footpoint, particularly for larger $\delta$ values. This is expected, as the separation between reconnecting field lines increases with distance from the reconnection site, as shown in Figure \ref{Cartoon}. Since field lines are traced from the lower boundary, a reconnection site closer to this boundary leads to a greater vertical divergence between pre- and post-reconnection field lines, resulting in a larger $\delta$.  In contrast, a reconnection site located close to the top boundary would produce a smaller $\delta$. This asymmetry provides indirect evidence that the measured $\delta$s are due to reconnection rather than numerical slippage of the field lines. Unlike reconnection, slippage would affect the field line more uniformly and would not preferentially align with regions of maximum viscous heating.  We discuss slippage further in the Appendix \ref{slip_recn}.

\cite{Knizhnik_2022ApJ...937...93K} found a different behavior. The number of reconnections decreased with height in the corona. Their simulation did not include the transition region and chromosphere, and used different driving patters (rotational cells at fixed locations) compared to ours. However, their magnetic configurations do account for field expansion with height. This is a realistic feature that we plan to incorporate into future simulations.

\subsubsection{Define a nanoflare }

\begin{figure*}
 \centering
 \includegraphics[width=0.95\textwidth,angle=0]{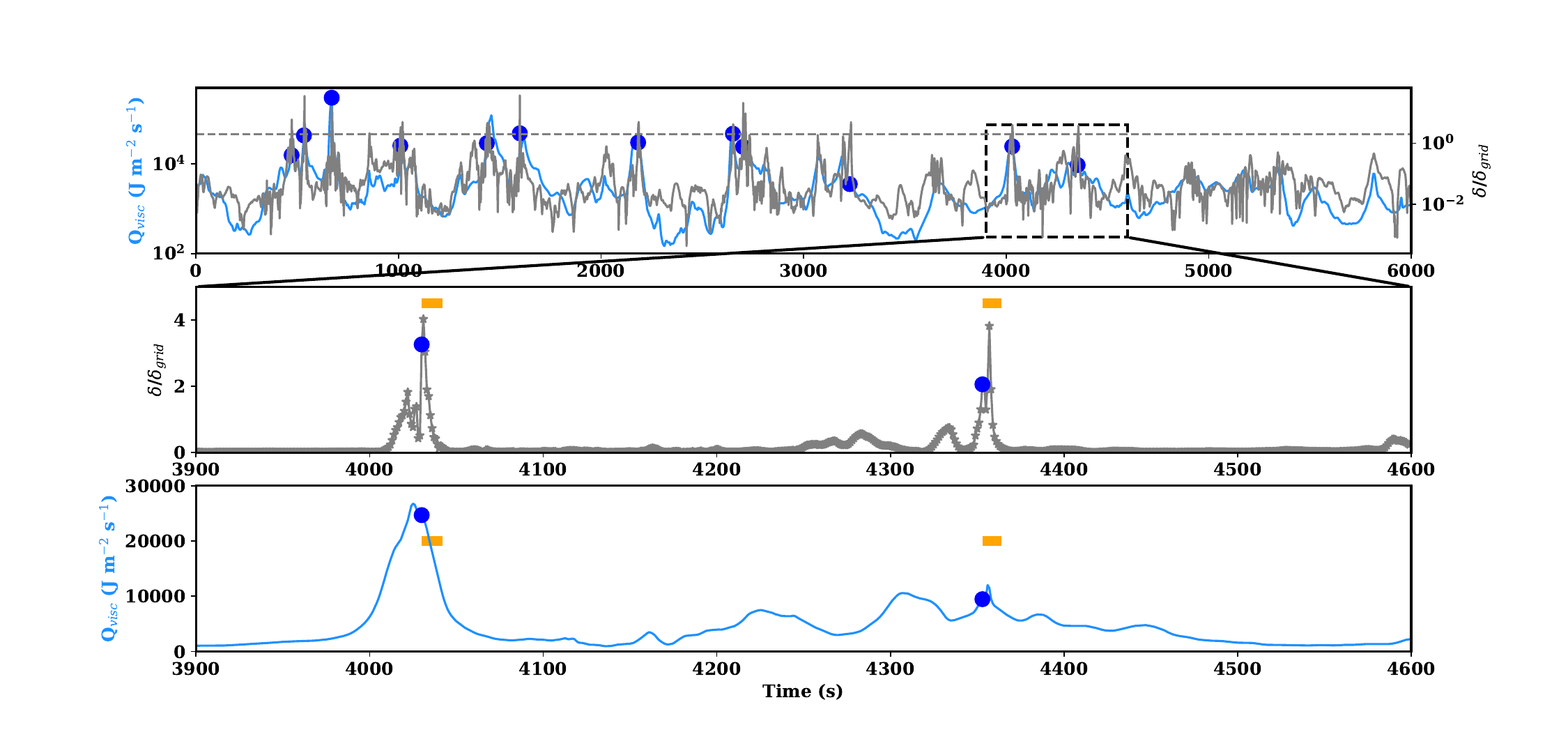}
 \caption{Top panel: Coronal integrated viscous heating rate and reconnection jumps normalized by the grid resolution. Blue circles are nanoflares identified from Method A.  Middle panel: Zoom-in of the reconnection jumps. The collective nature of the jumps demonstrates multiple reconnections while the field line crosses the current sheet diffusion region. The durations of the nanoflares are shown with orange bars for $t_{window}$ = 10 s. Bottom panel: Corresponding zoom-in of the viscous heating rate. Method A captures only a part of the energy released by the reconnections because the defined nanoflare duration fails to cover the entire heating event.}
  \label{methodA_caveat}
\end{figure*}

A key feature of this simulation is viscous shock heating. When reconnection occurs, the resulting reconnection jets generate shocks that compress the surrounding plasma. The high velocity gradient at the shock front leads to viscous heating, which the simulation identifies as a primary source of heating associated with reconnection. For each field line, we quantify the coronal integrated viscous heating rate ($Q_{visc}$), represented by the blue curve in the top panel of Figure \ref{delta_vs_time} for one randomly chosen field line. Here, a field line is identified by its footpoint at the lower driver plane, which is advected with the flow. There is a strong correlation between $Q_{visc}$ and the reconnection jumps ($\delta$) experienced by the field line, as indicated by the grey curve.

Unlike the conventional definition of a \textit{nanoflare} which describes it as an impulsive energy release resulting from all magnetic reconnection events within a flux bundle, our definition characterizes nanoflares as heating events experienced by a single field line. The observed nanoflares, in this case, could arise either from flipping reconnection within a single finite-width current sheet, where a field line reconnects multiple times in sequence as it passes through the sheet \citep{Priest_2003JGRA..108.1285P, Pontin_2005PhPl...12e2307P}, or from distinct reconnection events occurring in separate current sheets. In the conventional case, the energy of a nanoflare depends on the cross-sectional area of the entire flux bundle over which the energy is released. In contrast, we define nanoflare energy as the total energy gained by a field line through one or multiple reconnections involving single or multiple current sheets. All nanoflare quantities (e.g. nanoflare energy, occurrence frequency) measured by the methods described below are expressed per unit area or per field line.

\subsubsection{Method A}
A nanoflare starts when a coronal reconnection jump satisfies $\delta \ge \delta_{cut}$. This occurs when a grey point in the bottom panel of Figure \ref{delta_vs_time}, exceeds the grey dashed line which corresponds to $\delta_{cut} = 2$ grid cells. Simultaneously, the height of maximum viscous heating rate, represented by the blue point should lie between the blue horizontal lines which mark the coronal part of the field line. Although the reconnection connectivity change is an instantaneous process, the consequent conversion of magnetic energy into thermal energy occurs over a longer timescale. Plasma is first accelerated in the outflow jet and then decelerated as the jet encounters slower plasma in its way, with energy changing from magnetic to kinetic to thermal. We employ a waiting time of $t_{window}$ to account for this finite duration and define the end of the nanoflare to be the end of the waiting time if no other reconnection occurs. If another reconnection does occur during the waiting time, we extend the waiting time to $t_{window}$ after that reconnection. This continues until there is no additional reconnection during the last $t_{window}$. The duration of the nanoflare thus extends from the time of the first reconnection to $t_{window}$ after the last reconnection.
Nanoflares identified using this method are displayed in the bottom panel, where the orange bars represent the durations of the identified nanoflares with $t_{window}$ = 10 s in this case. 
\begin{figure*}
 \centering
 \includegraphics[width=0.95\textwidth,angle=0]{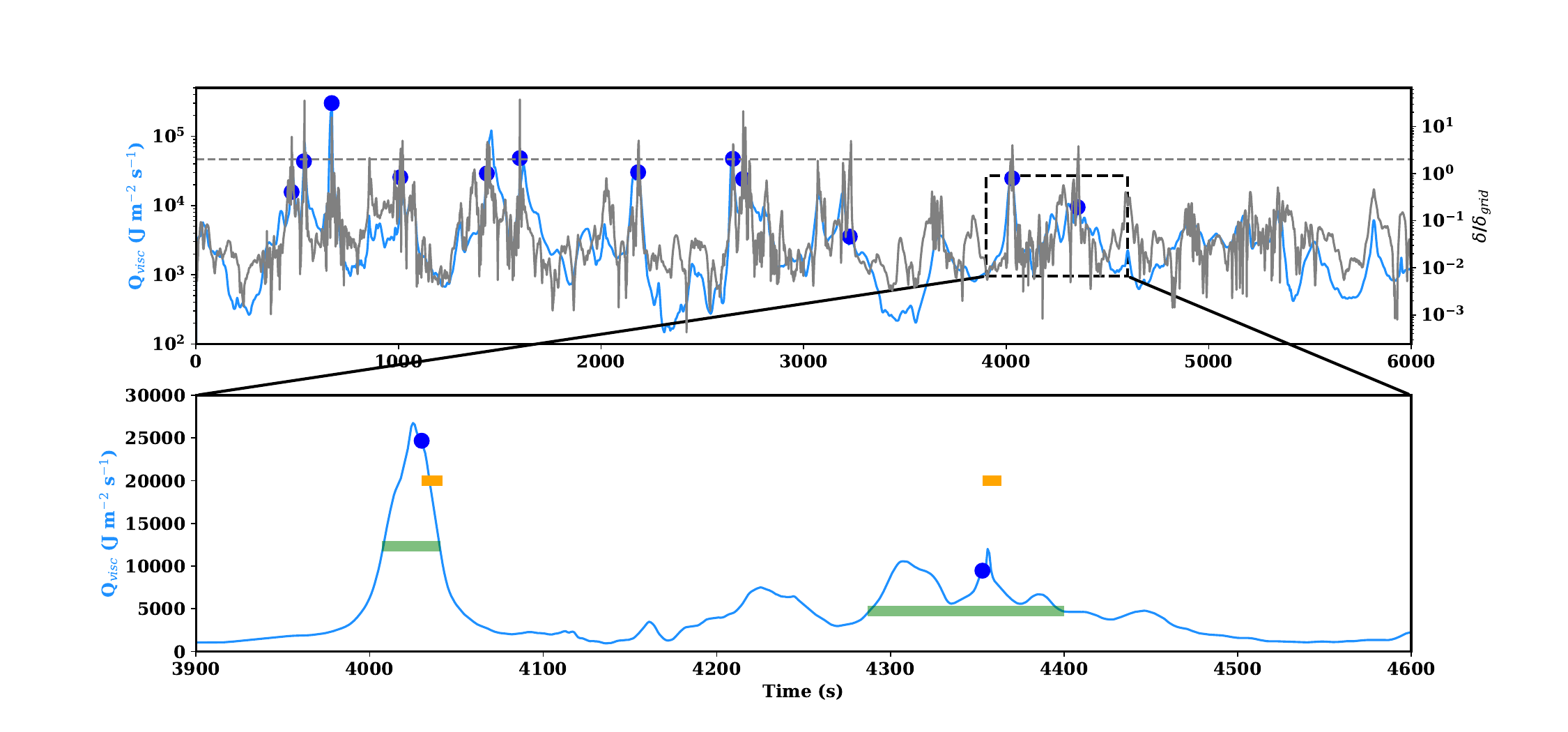}
 \caption{A comparison plot showing nanoflares identified from Methods A and B. The orange and green bars are the nanoflare durations from Methods A and B, respectively. Blue circles are the nanoflares identified from both the methods. $t_{window}$ used in this case is 10 s.}
  \label{methodAB}
\end{figure*}

Figure \ref{methodA_caveat} is an another example which shows the viscous heating rate for the same field line shown above but for a different zoomed in time range. The blue circles mark the starts of the nanoflares identified from the method described above. The middle panel shows two more examples where a single nanoflare involves multiple $\delta$ reconnection jumps. These are not distinct events at different locations along the field line. Rather, the field line experiences continuous reconnections as it diffuses across the diffusion region at a single current sheet location \citep{Priest_2003JGRA..108.1285P, Pontin_2005PhPl...12e2307P}. The magnetic field direction rotates continuously across the current sheet - it is not a discontinuity - so a fixed footpoint at one photospheric boundary maps to what appears to be a rapidly moving footpoint at the other boundary. The plasma at that conjugate footpoint does not actually move. Rather, it is an illusion produced by the continuously changing connectivity. This manifests as grouping of reconnection jumps in the $\delta$ plot. We have verified that the time taken by a field line to cross the diffusion region is comparable to the time over which the clustering is identified, as expected.

It is quite evident from the bottom panel of Figure \ref{methodA_caveat} that this method has some weaknesses as it does not always capture the entire heating event. For example, each nanoflare duration (shown in orange bars, with $t_{window}$ = 10 s) encompasses only a small fraction of the released energy. In fact, the peak in the heating rate occurs even before the field has reconnected in this case. This kind of situation may arise when the field line experiences heating from outflows associated with an entirely different reconnection event involving different field lines. This can also be due to the constraint on the $\delta$ condition that we have imposed to define a reconnection event where small non-zero $\delta$s are not considered as reconnections.

Below we define another approach to define the duration of the nanoflare so that it can capture most of the energy released by the identified reconnections.
\begin{figure*}
 \centering
 \includegraphics[width=0.95\textwidth,angle=0]{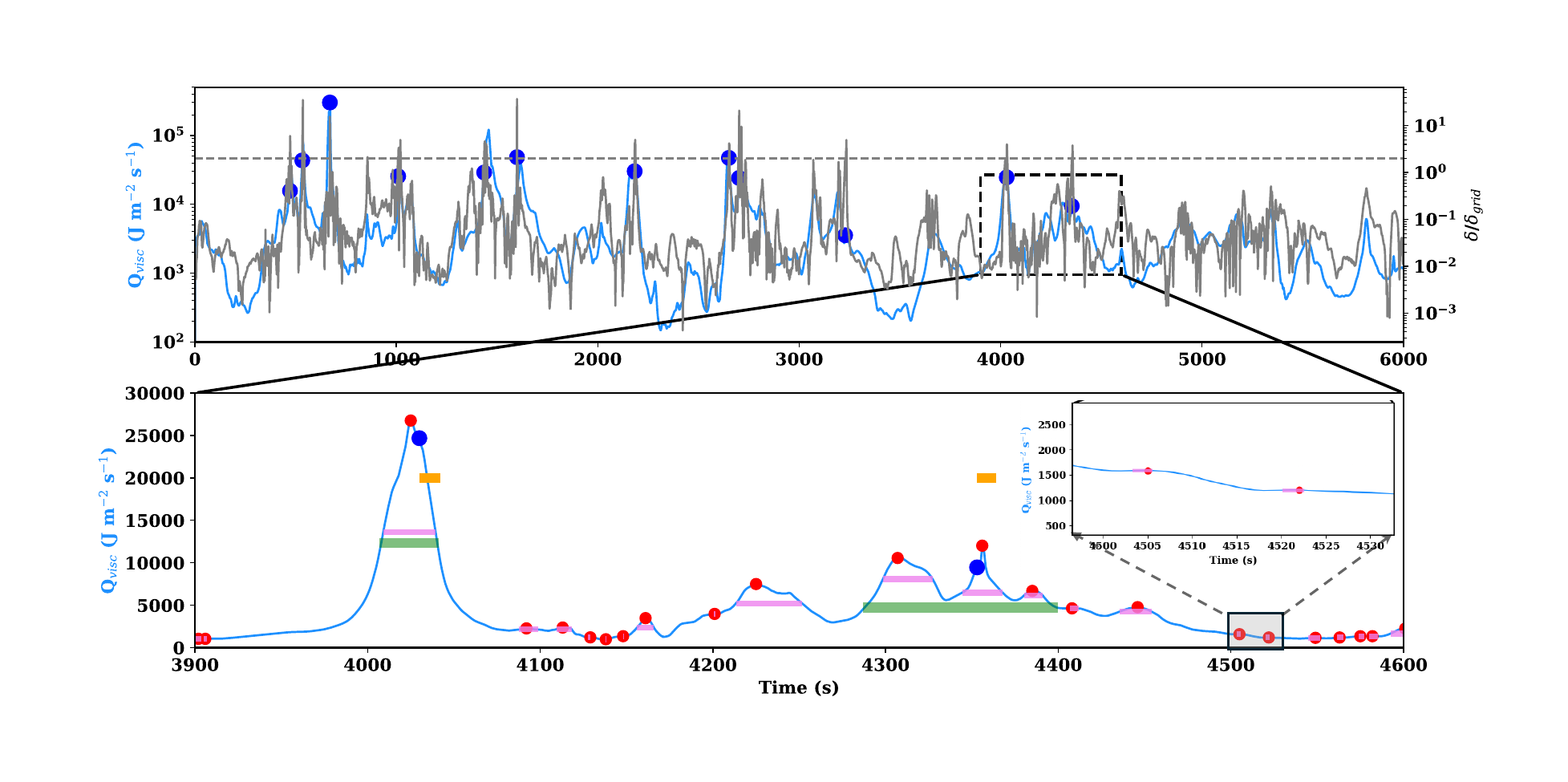}
 \caption{A comparison plot showing nanoflares identified from Methods A, B and C. The orange, green and purple bars are the nanoflare durations from Methods A, B and C, respectively with $t_{window}$ = 10 s. The zoomed in box in the bottom panel shows the presence of short duration nanoflares not visible in the original panel. Blue circles are nanoflares from Methods A and B, while nanoflares from Method C are marked by the red circles.}
  \label{methodABC}
\end{figure*}

\begin{figure}
 \centering
 \includegraphics[width=0.45\textwidth,angle=0]{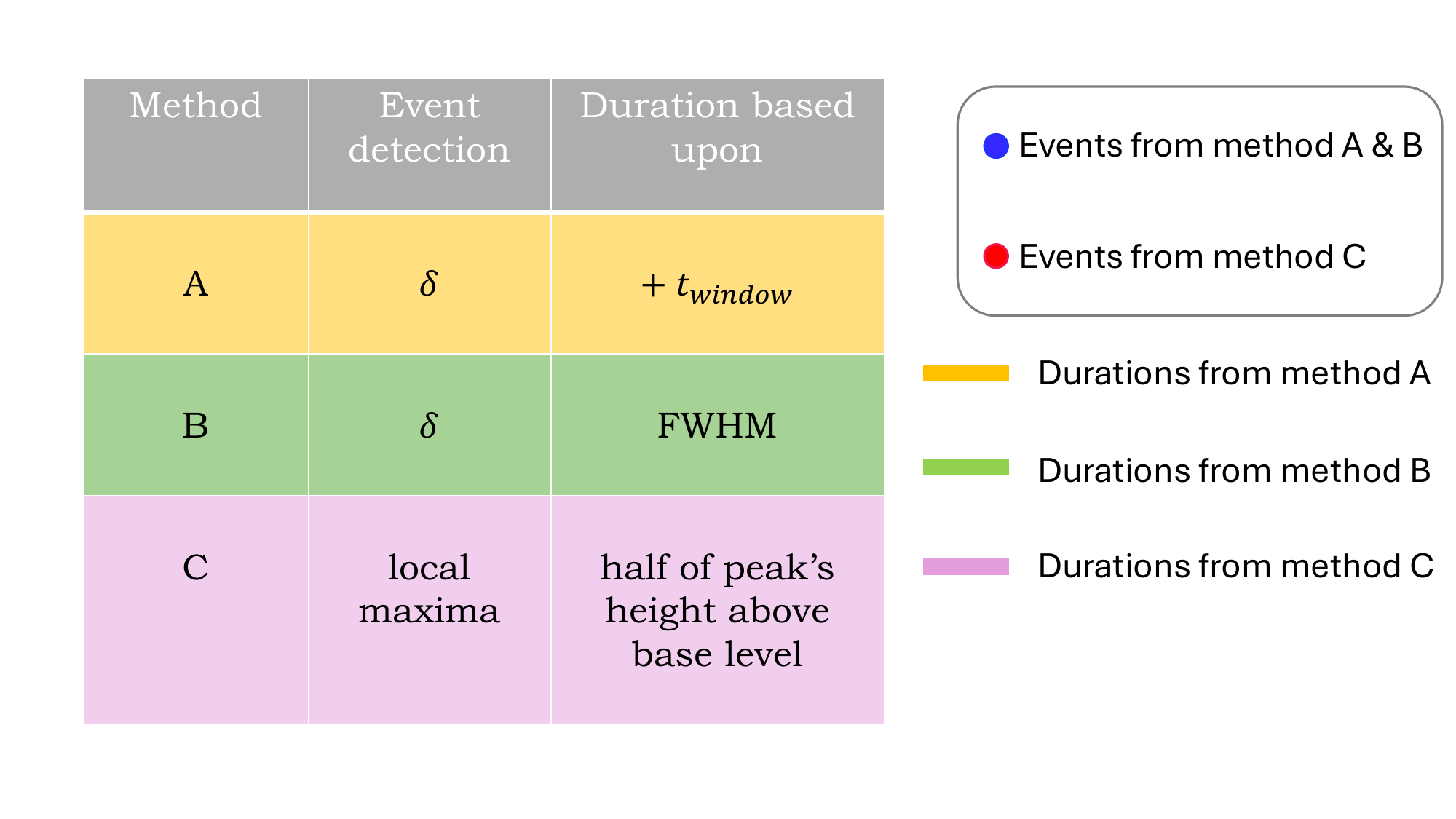}
 \caption{Table summarizing all the three methods introduced in this paper to define a nanoflare and its duration and energy.}
  \label{tab:diff_methods}
\end{figure}

\begin{figure*}
 \centering
 \includegraphics[width=0.95\textwidth,angle=0,page=1]{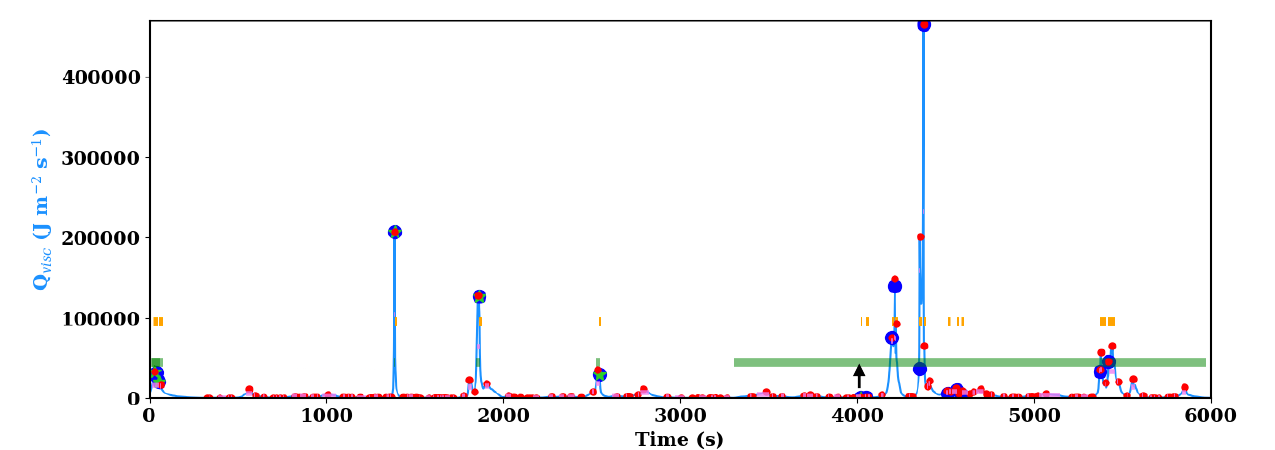 }
 \caption{Evolution of coronal integrated viscous heating rate for a randomly chosen field line over the entire 6000 s tracking period. Blue circles, green stars, and red circles indicate the times when nanoflares are identified from Methods A, B, and C, respectively with $t_{window}$ = 10 s. The corresponding durations are indicated by orange, green, and purple bars. The nanoflare identified by the black arrow near 4000 s is an example of how Method B can sometimes measure extreme durations ($\sim 2500$ s in this case.)}
  \label{long_duration}
\end{figure*}

\begin{figure*}
 \centering
 \includegraphics[width=0.95\textwidth,angle=0]{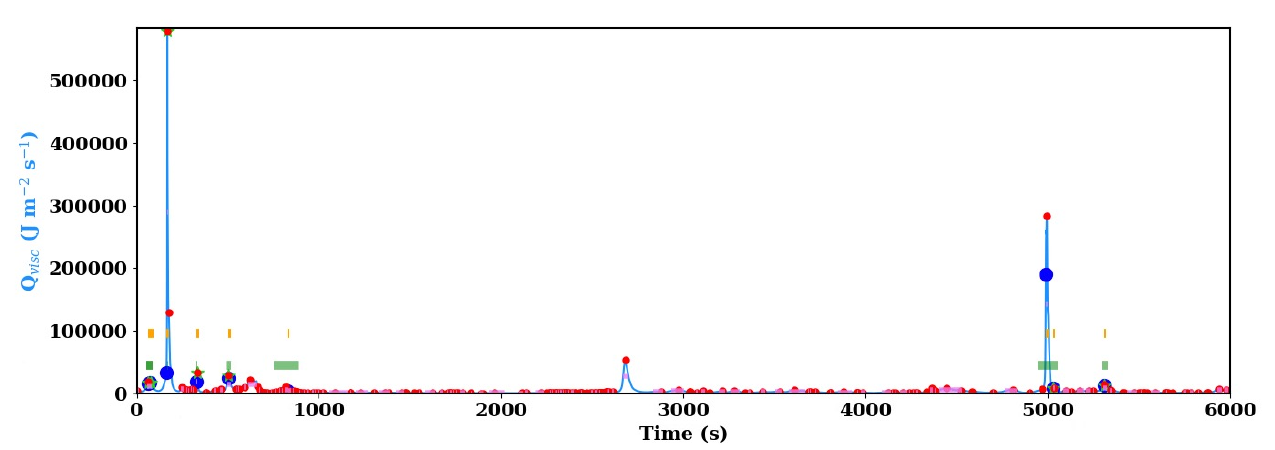 }
 \caption{An example when the delay between consecutive nanoflares can be as long as 4000 s for Methods A and B. }
  \label{long_delay}
\end{figure*}

\subsection{Method B}
The nanoflares in this method are identified using the previous ($\delta$) method, but the duration is taken to be the full-width at half maximum (FWHM) of the viscous heating profile, where the `maximum' is the strongest coronal integrated viscous heating rate within the previously estimated duration of the same nanoflare from Method A. Figure \ref{methodAB} is a comparison plot, where the green bars are the durations from this FWHM method and the orange bars are from Method A with $t_{window}$ = 10 s. It is evident that this method captures the viscous heating that was missed by Method A. It also tends to aggregate temporally adjacent events of comparable heating rates.

However, this FWHM method is still unable to capture the remaining, generally small, peaks in the heating rate curve that do not satisfy the $\delta \ge \delta_{cut}$ condition. It could be because the condition is too severe to identify small jumps of the field line or because the heating is due to other nearby reconnections not directly involving this field line. As a result, we present a third technique which captures all of the heating peaks without making a distinction as to whether or not the field line has reconnected.

\subsection{Method C}
In this method, a nanoflare is defined whenever there is a peak in the coronal integrated viscous heating rate for the tracked field line. To identify the peaks we have used the python routine \texttt{find\textunderscore peaks} which uses local maxima for peak detection. Once a peak is identified, a nanoflare is defined as follows. First, a background level is set that is equal to the largest local minimum on either side of the peak. The duration of a nanoflare is the interval over which the heating rate exceeds the background by at least half the amount that the peak exceeds the background. The python routine  \texttt{peak\textunderscore widths} is used to make this determination. Often events overlap, in which case we merge them into a single event. Note that the energy of the event is relative to zero, not the background level, i.e., we do not subtract the background.

\begin{figure*}
  \centering
      \includegraphics[width=0.95\textwidth,angle=0]{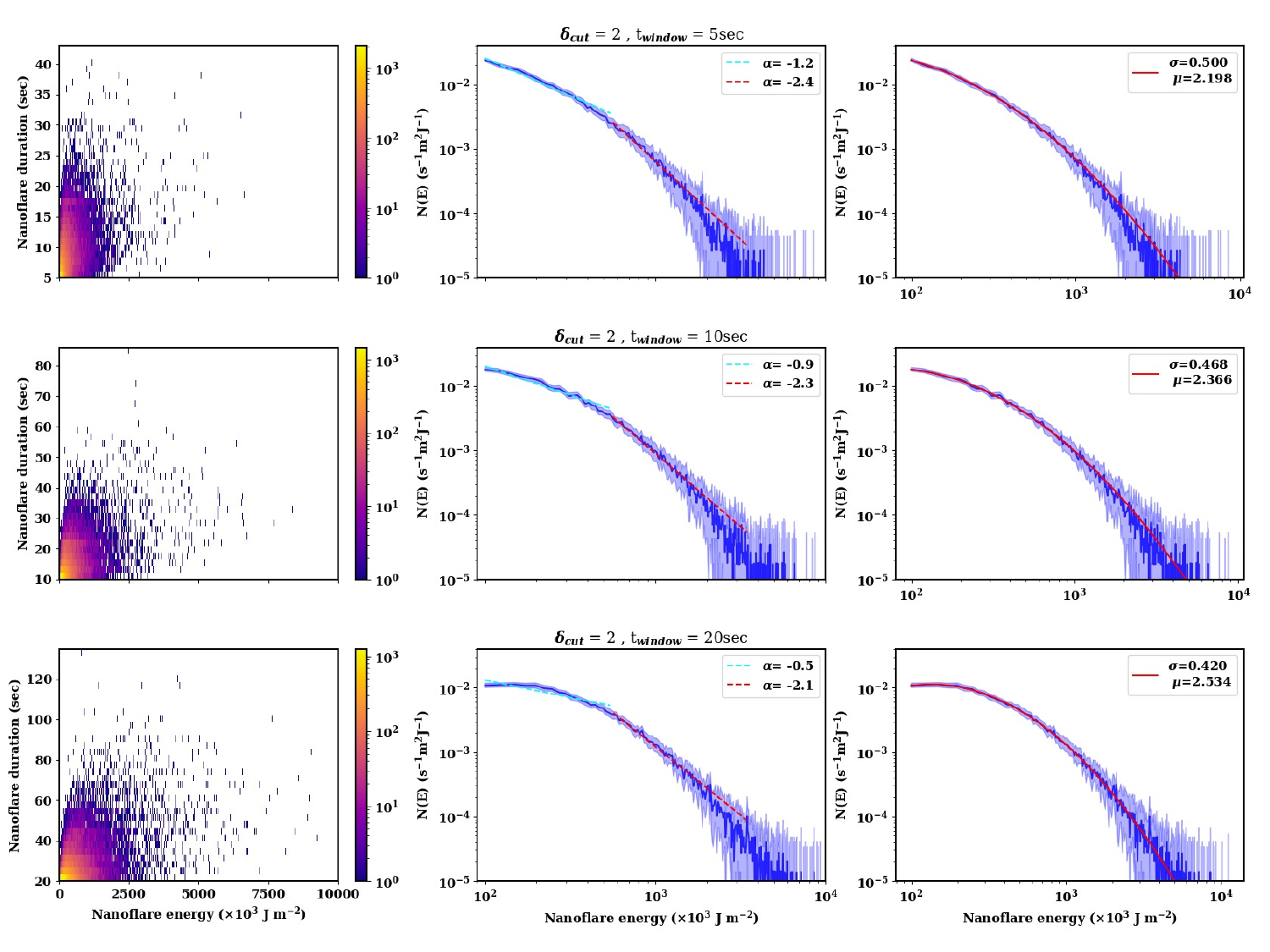}
 \caption{Nanoflare energies and durations from Method A with $\delta_{cut}$=2 grid cells and $t_{window}$=5, 10, and 20 s (top to bottom). Left column shows the 2D histograms of nanoflare energies and durations. Middle and right columns show the energy distributions obtained by collapsing the histograms in the vertical direction. Double power-law fits and log-normal fits are indicated in the middle and right columns, respectively.}
  \label{methodA}
\end{figure*}

\begin{figure*}
  \centering
      \includegraphics[width=0.95\textwidth,angle=0]{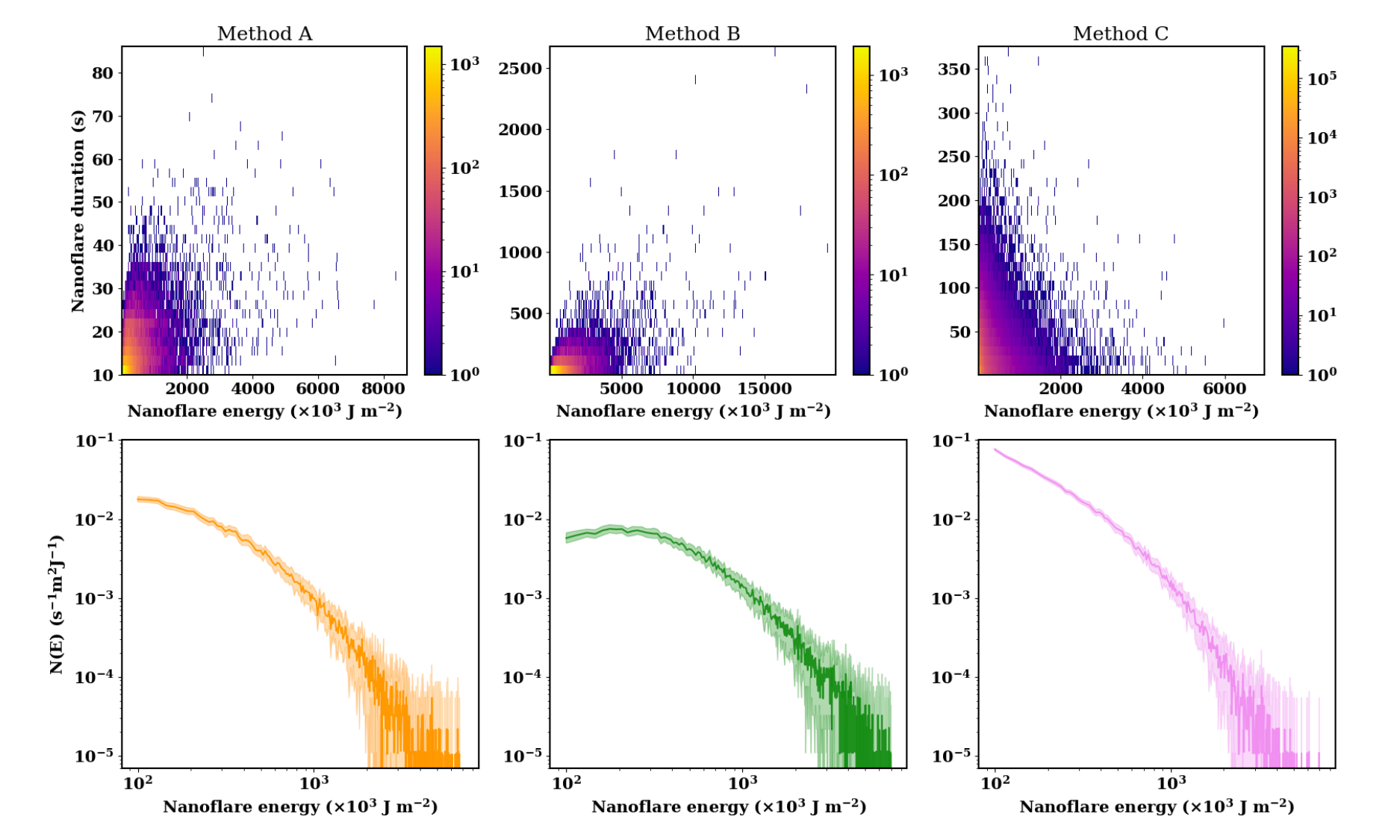}
 \caption{Comparison between Methods A, B and C with $\delta_{cut}$=2 grid cells and $t_{window}$=10 s. (Top panel) 2D histograms of nanoflare energies and durations. (Bottom panel) Nanoflare energy distributions.}
  \label{three_methods_histograms}
  \end{figure*}

Figure \ref{methodABC} is a comparison plot which shows nanoflares and their corresponding durations as identified by all three methods described above. The blue circles are the nanoflares detected by the $\delta$ method and red are the nanoflares from local maxima method. The durations of the smaller events are difficult to see in the plot, but are more clear in the zoomed inset for two events. It is to be noticed that the third method detects a significantly greater number of events compared to the $\delta$ method. Methods A \& B capture nanoflares only when the $\delta$ condition is satisfied by the coronal reconnections, which means reconnection jumps smaller than $\delta_{cut}$ will be missed, thereby reducing the number of cases in these two methods compared to Method C. 

Figure \ref{tab:diff_methods} summarizes all the three methods described above. Figure \ref{long_duration} shows an example that compares different methods for a randomly chosen field line. It shows the corona-integrated viscous heating rate for a single field line and for the entire simulation time. Blue circles, green stars, and red circles indicate nanoflares as identified from Methods A, B, and C, respectively. This example includes an extreme case showing that a nanoflare duration can be as long as 2500 s for Method B (shown by the green bar). This can happen when the method identifies a nanoflare with very small heating rate. In that case, the half maximum lies very close to the background heating rate which makes the FWHM much longer than any event with stronger heating rate. Method C detects a much greater number of events, and their duration is very small compared to the events detected from Methods A and B. Figure \ref{long_delay} shows an example when the delay between consecutive nanoflares can be as long as 4000 s for Methods A and B. These are some extreme cases showcasing the comparison between all the three methods. 

We performed this analysis for 3136 field lines. We tracked the field lines for 6000 s and identified all the nanoflares that occurred using the three different methods. We determined the energies of the nanoflares by spatially integrating viscous heating rate over the coronal part of the field line - as done for the plots - and temporally integrating over the duration of the event. However, it is to be noted that the combined energy of all the nanoflares is only a fraction of total viscous heating because heating also occurs outside the times of the nanoflares. A more detailed discussion on the fraction of the total energy contributed by nanoflares, as captured by the different methods, is provided in the Appendix \ref{energy_fraction}. 

\section{Results}\label{sec:Results}

Figure \ref{methodA} (left column) shows 2D histograms of nanoflare energy and duration as determined from Method A with $\delta_{cut} = 2$ grid cells and $t_{window} = 5,10,$ and $20$ s. Color indicates the number of nanoflare occurrences of given energy and duration. The energy is expressed as energy per unit area. This differs from conventional flare energy, which is typically volume-integrated rather than calculated per field line or unit area. We make no attempt here to measure the total energy of spatially large nanoflare events that encompasses multiple adjacent numerical cells. 
Our ultimate interest is in how plasma responds to impulsive heating and produces a spectrum of emitted radiation. Once the heating has occurred, this is controlled by field-aligned physics via processes like thermal conduction, chromospheric evaporation, and draining of cooling plasma. It is the heating per unit area that matters in this regard. The middle column shows the energy distribution of nanoflares where the $y-$axis indicates the number of nanoflares per unit time per unit energy. This is obtained by summing the events over all durations at a given energy bin (collapsing the histogram along the y-axis) and normalizing the resultant energy distribution with the bin size ($\Delta E$) and total tracking time of the field lines (= 6000 s).

To take into account the variations in nanoflare energy distributions that one might anticipate if a different set of initial footpoints had been selected, we have applied the Bootstrap approach. With Bootstrap, one can compute the sample statistics each time by repeatedly resampling the original sample values, with replacements. The range that the energy distribution can fluctuate over, as obtained by bootstrapping the original sample, is displayed by the light blue band. 

As evident from the middle column, none of the distributions can be fitted by a single power-law. We divide the energy range into two parts and fit power-law functions whose exponents are shown in legends of the respective plots. Clearly, the exponent is consistently flatter than -2 in the lower energy range and steeper than -2 in the higher energy range. For single power-law distributions, low energy events dominate the total energy for slopes steeper than -2, and high energy events dominate the total energy for slopes flatter than -2. This is the opposite of what we find, indicating that neither low nor high energy nanoflares dominate the heating in our simulation. Rather, most of the heating is produced by nanoflares of intermediate energy range. These mid-size events are well measured, as indicated by the small `error bars' of the bootstrap analysis. The fact that the slope is flatter than -2 at low energies is especially significant. It means that events too small for us to detect ($< 10^5$ J m$^{-2}$) are energetically unimportant.

The third column of Figure \ref{methodA} shows that the nanoflare energy distributions are, in fact, well fitted with a log-normal function:

\begin{equation}
N(E) = \frac{N_{0}}{E\sigma \sqrt{2\pi}} e^{-{\frac{1}{2}(\frac{ln(E)-\mu}{\sigma})^2}} , 
\end{equation}

\noindent where $N_{0}$ is the amplitude, $\mu$ is the mean, and $\sigma$ is the standard deviation of the distribution. The parameters in the legends are from the best fit obtained by minimizing the root mean square difference between the original and fitted values. 

There can be multiple reasons for deviation from a single power-law distribution which is typically observed for full-size flares. First, the flare energies are integrated over the entire flaring volume, whereas we use the energy per unit area for reasons discussed earlier. Second, flares may be fundamentally different from nanoflares. Eruptive flares in particular have completely different magnetic morphologies. Nanoflares are closer to confined flares, and it is interesting that flares without coronal mass ejections have a power-law exponent steeper than -2, whereas flares with CMEs have an exponent flatter than -2 \citep{Yashiro_2006ApJ...650L.143Y}. In addition, a comprehensive study of 6924 solar flares detected by SDO/AIA and 9601 flares detected by GOES/XRS between 2010 and 2018 reveals that most background-corrected flare data are better described by a log-normal distribution than a power-law \citep{Verbeeck_2019ApJ...884...50V}. However, raw GOES/XRS peak flux fits a power-law. These findings challenge the assumption of power-law behavior in solar flares and suggest that its implications, such as for coronal heating, require careful reevaluation. The authors hypothesize that the observed distribution results from the constant fragmentation of magnetic flux elements in the vicinity of loop footpoints, a process that \cite{Bogdan_1988ApJ...327..451B} has shown would result in a log-normal distribution. The magnetic field in our simulation also experiences constant fragmentation from a combination of complex photospheric driving and coronal reconnections \citep{Klimchuk_2021FrASS...8...83K}. 

\cite{Kalman_2018ApJ...853...82K} studied the response of the coronal field to imposed photospheric driving and used three different proxies to identify nanoflare events:  temperature changes ($\Delta T$), horizontal Poynting flux ($S_{h}$), and current density ($|\bm{J}|$), all in the mid-plane. Event ``energies" were obtained by integrating the proxy over the entire spatiotemporal domain of the event, where the event area and lifetime were determined using the cluster identification method \citep{Vadim_2010JGRA..115.9205U, Vadim_2010PhRvE..82e6326U}. They fitted the energy distributions with a single power law and found slopes of -1.76, -2.10, and -1.56 for $\Delta T$, $S_{h}$, and $|\bm{J}|$, respectively. However, the shapes of the distributions have log-normal aspects. It should also be noted that the event energies are area integrated and not per unit area, so the statistics are more like full-size flare statistics. 

It is important to note that the log-normal parameters $\mu$ and $\sigma$ may vary depending on the simulation setup, including factors like the initial length of the field lines and driving pattern. Future studies should investigate how these parameters change under different conditions.

We have verified that the results presented in Figure \ref{methodA} are largely unchanged with $\delta_{cut} = 1$ and $3$ grid cells. The distributions and fit parameters are very similar.

\begin{figure*}
 \centering
 \includegraphics[width=0.95\textwidth,angle=0]{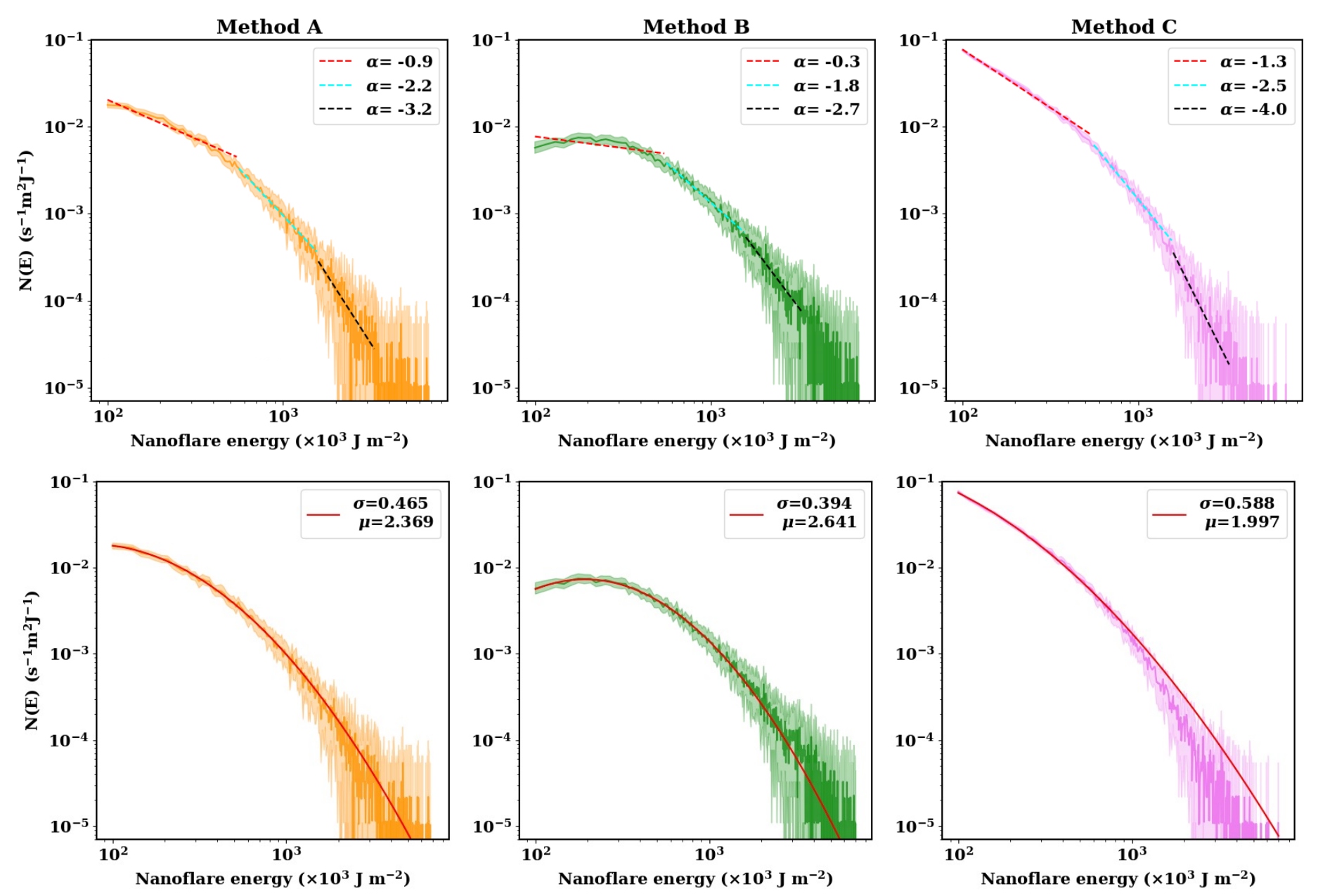}
 \caption{Power-law (top) and log-normal (bottom) fits of the nanoflare energy distributions obtained from Methods A, B and C with $\delta_{cut}$=2 grid cells and $t_{window}$=10 s.}
  \label{three_methods_fitting}
  \end{figure*}

We now compare the results for Methods A, B, and C where $\delta_{cut} = 2$ grid cells and $t_{window} = 10$ s. 
Figure \ref{three_methods_histograms} shows 2D histograms of nanoflare energies and durations (top) and energy distributions (bottom) for the three methods. The event durations from Method A are generally much shorter than those from Methods B and C, with most values in the range 10-30 s and a maximum value far less than for the other two. The energies from Method A are correspondingly weaker than from Method B, as clearly evident in the histograms. As previously described, Methods A and B only capture heating events in which the field line is directly involved in the reconnection. As a result, the number of events is far less than with Method C, which also captures events when the field line does not itself reconnect. These additional events tend to be of lower energies. They produce a marked increase at the low energy end of the energy distribution compared to Methods A and B. The local slope nonetheless remains shallower than -2. Method B shows some extreme cases where the durations can be as long as 2500 s, an example of which is shown in Figure \ref{long_duration}. 

The top and bottom panels of Figure \ref{three_methods_fitting} show power-law and log-normal fits of the nanoflare energy distributions from Methods A, B and C. As before, none of the distributions are well fitted by a single power-law, but well fitted by log-normal. For all three methods, the power-law slopes are much flatter than -2 in the lower energy regimes and steeper than -2 in higher energy regimes implying that mid-energy nanoflares contribute more significantly to the total energy budget. The light orange, green, and pink bands represent the range of energy distribution fluctuations obtained by bootstrapping the original sample for Methods A, B, and C, respectively, while the dark color indicates the mean of the generated distributions.
 
Nanoflare frequencies can be classified into two categories.
\begin{enumerate}
    \item [(I)] \textit{Energy-dependent repetition frequency:} Frequency with which nanoflares in a given energy range repeat. This is demonstrated by the above energy distributions (Figure \ref{three_methods_fitting}), which show that low energy nanoflares are more numerous and therefore repeat more often than high energy ones. We present more detailed analysis of nanoflare repetition frequency per field line, which supports this interpretation, in Appendix \ref{nanoflares_per_fl}.
    
    \item [(II)] \textit{Energy-independent repetition frequency:} This is the repetition frequency of any two successive nanoflares independent of their energies. This reflects the delay between two consecutive nanoflares of any energy. The energy-independent frequency is the most commonly discussed in the literature.
\end{enumerate}

\begin{figure*}
 \centering
 \includegraphics[width=0.95\textwidth,angle=0]{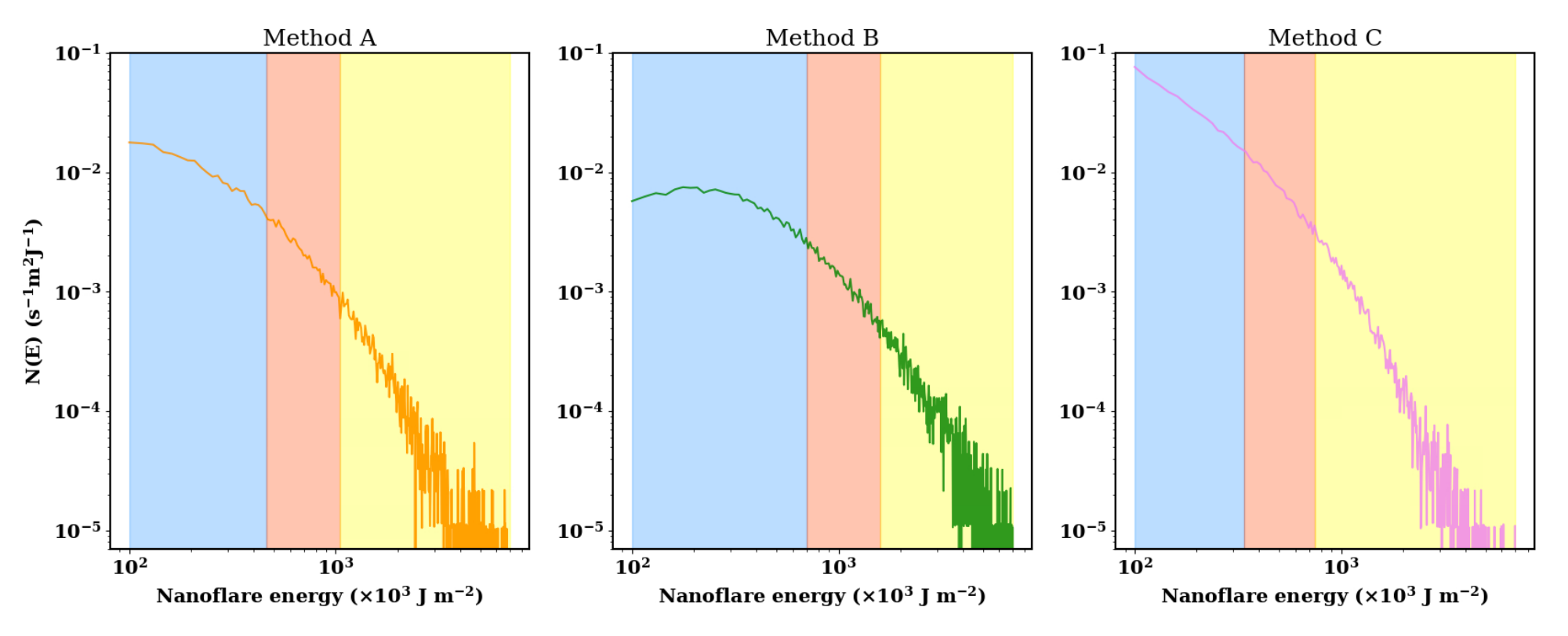}
 \caption{Energy partition of nanoflares. Blue, red and yellow columns -- each contains nanoflares that constitute 1/3 of the total energy and  are classified as low, medium and high energy events. The distributions are the same as in Figures 13 and 14.}
  \label{occurrence_frequency_three_energy_ranges}
\end{figure*}

\begin{figure*}
 \centering
  \includegraphics[scale=0.7,angle=0]{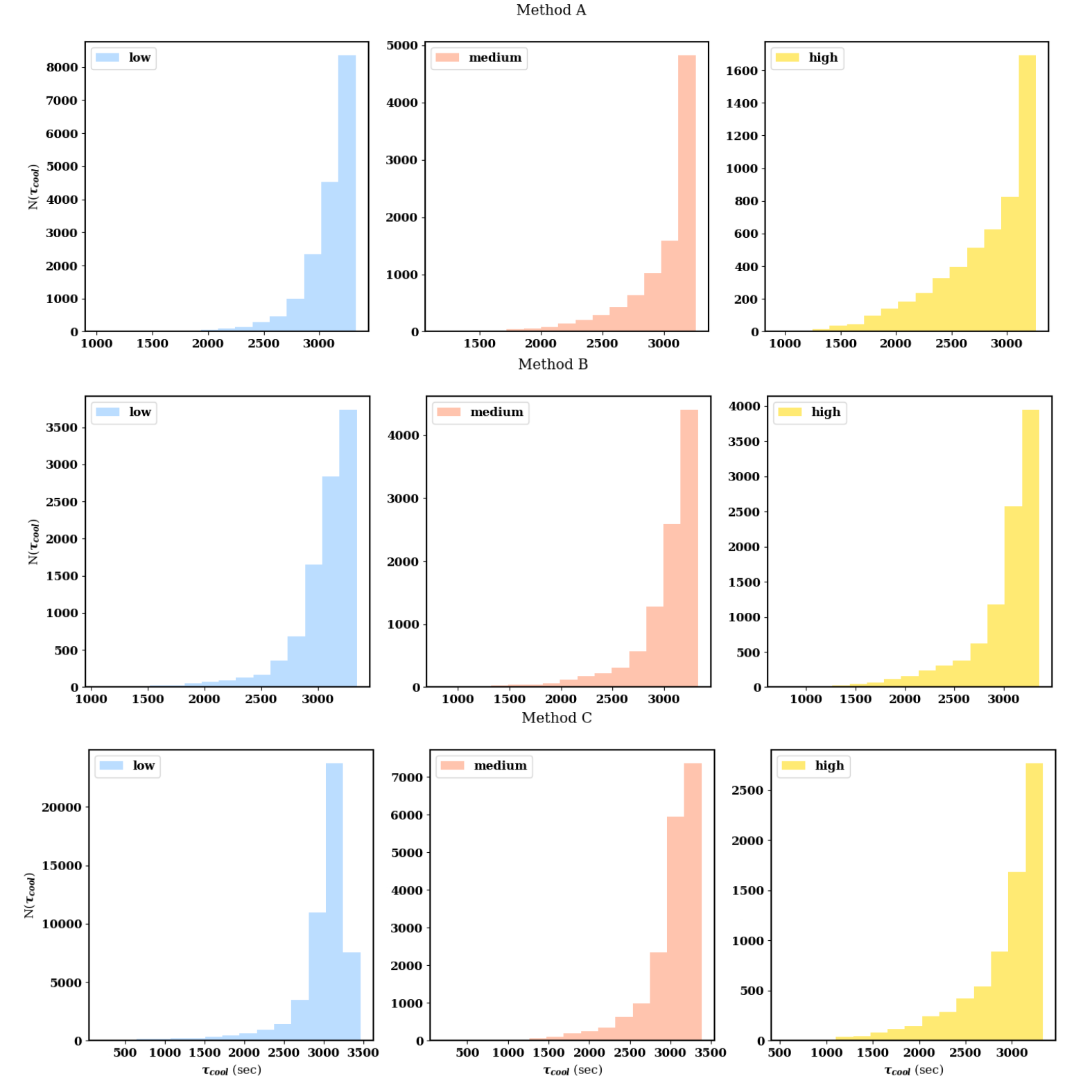}
 \caption{Histograms of cooling time for low, medium and high energy nanoflares represented in blue, red and yellow, respectively.}
  \label{cooling_time_histogram}
\end{figure*}

\begin{figure*}
 \centering
 \includegraphics[width=0.95\textwidth,angle=0]{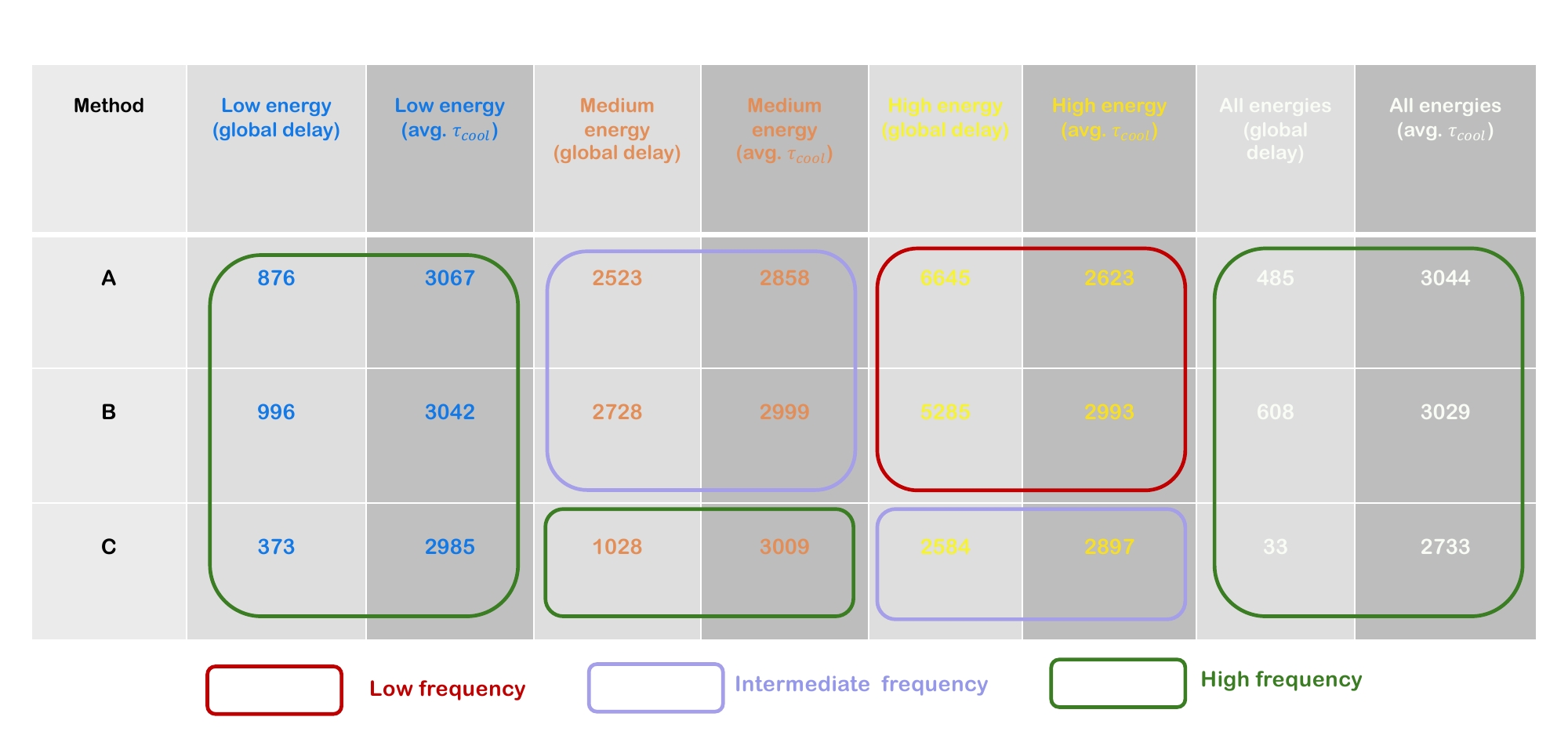}
 \caption{Table summarizing the global delays and average cooling times in seconds for different energy nanoflares.}
  \label{tab:summary_table}
\end{figure*}

\begin{figure*}
 \centering
 \includegraphics[width=0.95\textwidth,angle=0]{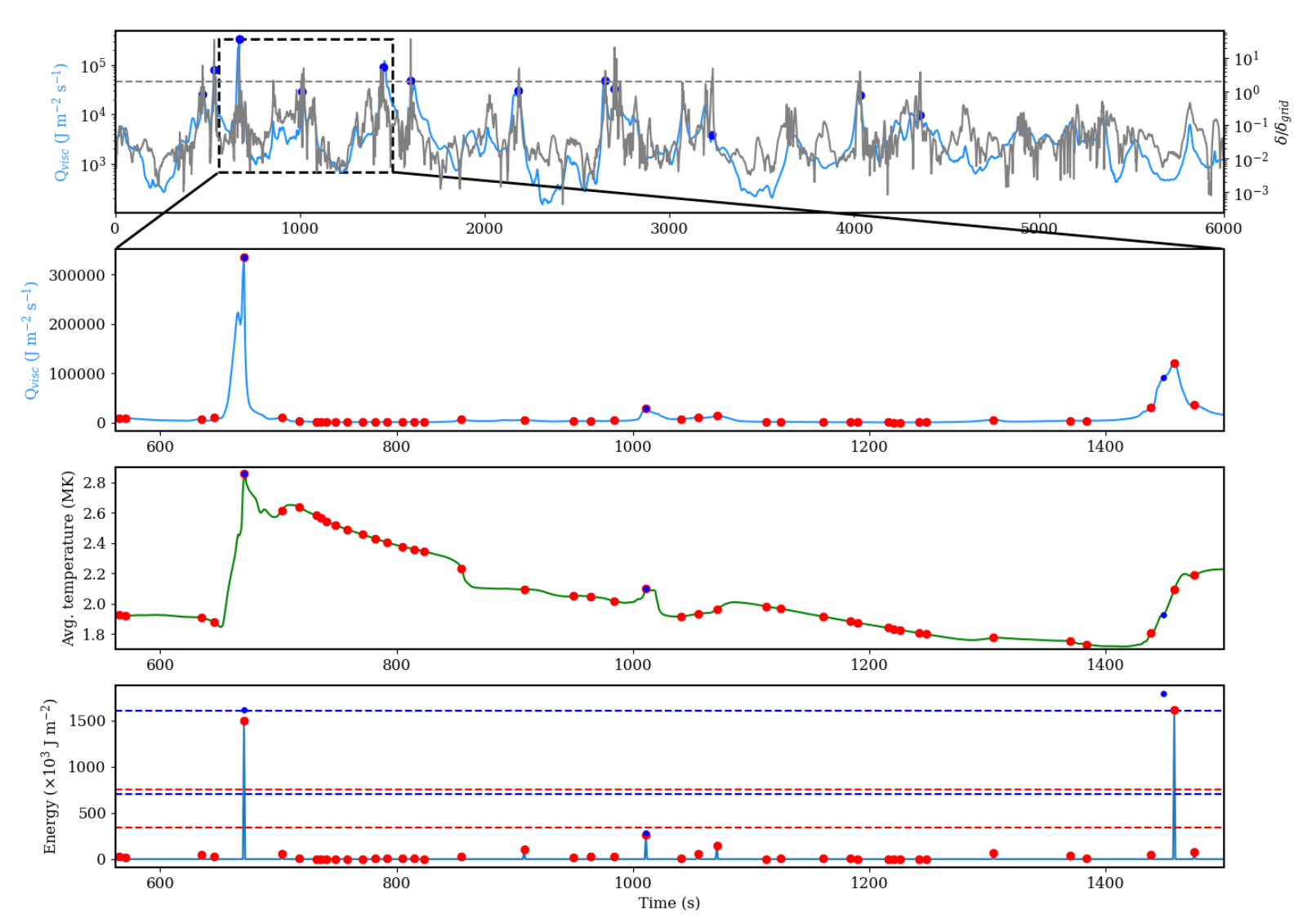}
 \caption{An example showing the role of different energy nanoflares in determining the overall temperature evolution.}
 \label{example_high_energy}
 \end{figure*}

Nanoflares in the simulation are classified into three energy ranges—low, medium, and high—each containing one-third of the total energy. This enables calculation of delays between successive nanoflares of similar energy. Figure \ref{occurrence_frequency_three_energy_ranges} illustrates the energy distribution, with blue, red, and yellow columns representing the energy ranges. The global delay between successive nanoflares in each energy range is calculated from the following.

\begin{equation}
    Global \, \, delay, \mathcal{D} = \frac{1}{\mathcal{R}}
\end{equation}
where 
\begin{equation}
    Global \, \, rate, \mathcal{R} = \frac{\mathcal{N}_{tot}}{\tau \times \mathcal{N}_{pts}}
\end{equation}
Here, $\mathcal{N}_{tot}$ is the total number of nanoflares in a particular energy range (i.e., in either the blue, red or yellow columns), $\tau$ (= 6000 s) is the total time over which the field lines are tracked and $\mathcal{N}_{pts}$ (= 3136) is the total number of footpoints tracked in this study. When $\mathcal{N}_{tot}$  includes all the events in the entire energy range (low + medium + high energy), the computed delay is the average delay between successive nanoflares independent of their energies.

Nanoflare frequency is classified as low, intermediate, or high based on the delay between nanoflares relative to the plasma cooling time. Delays significantly longer than the cooling time are called low-frequency nanoflares, while delays comparable to and significantly shorter than the cooling time are called intermediate and high-frequency nanoflares, respectively. The plasma cooling time, which includes both conductive and radiative cooling \citep{Cargill_1995ApJ...439.1034C}, is calculated as follows. 

\textit{Conductive cooling time},
\begin{equation}
    \tau_c = \frac{21}{2}\Bigl(\frac{k_B n L^2}{\kappa_{0}T^{5/2}}\Bigr)
\end{equation}

\textit{Radiative cooling time},
\begin{equation}
    \tau_r = \frac{3k_B T}{n\Lambda(T)}
\end{equation}

\textit{Total cooling time},
\begin{equation}
    \tau_{cool} = \Bigl(\frac{1}{\tau_c}+\frac{1}{\tau_r}\Bigr)^{-1}
\end{equation}

 \noindent where $k_B$ is the Boltzmann constant, $\kappa_0=1.0 \times 10^{-6}$ erg s$^{-1}$ K$^{-1}$ cm$^{-1}$ is the coefficient of thermal conductivity along the field lines , $n$, $T$, and $L$ are the electron number density, temperature and half-length of the loop ($\sim 50$ Mm), respectively. $\Lambda(T)$ is optically thin radiative loss function adopted from \cite{Klimchuk_2008ApJ...682.1351K}. For each nanoflare, the plasma cooling time is estimated using the spatiotemporal average of $n$ and $T$. The spatial averaging is done over the coronal segment of the field line, and temporal average is over the corresponding nanoflare duration.

Histograms of cooling time determined for low, medium, and high energy nanoflares are shown in Figure \ref{cooling_time_histogram}. Top, middle, and bottom panels indicate nanoflares identified using Methods A, B, and C, respectively. Different colors indicate nanoflares with different energy ranges. Regardless of the energy range, all of the histograms indicate an average cooling time of $\sim$ 3000 s. To compare the cooling times with global delays between nanoflares in different energy ranges, we estimate the average cooling time ($\Bar{\tau}_{cool}$) from the histograms, which is given by, 
\begin{equation}
   \Bar{\tau}_{cool} = \frac{\sum N(\tau_{cool})\times \tau_{cool}}{\sum N(\tau_{cool})}
\end{equation}

\noindent where $N(\tau_{cool})$  is the distribution of total cooling times. 

Table \ref{tab:summary_table} summarizes the global delays and average cooling times for different energy nanoflares defined from the three different methods. Method A and B show a similar trend: low energy nanoflares occur at high frequencies (global delay $<$ average cooling time), medium energy nanoflares occur at intermediate frequencies (global delay $\sim$ average cooling time), and high energy nanoflares occur at low frequencies (global delay $>$ average cooling time). In contrast, nanoflares obtained from Method C occur within intermediate and high frequencies. This suggests that the active region exhibits a broad range of frequencies in which nanoflares with different energies repeat with different frequencies. In contrast, the last two panels suggest that the global delay between any two successive events (regardless of their energies) is significantly shorter than the average cooling time. This is primarily because low-energy events, which dominate in number, tend to repeat at high frequencies.
 
Figure \ref{example_high_energy} shows an example which explains the role of different energy nanoflares in determining the overall temperature evolution. The top panel shows the coronal integrated viscous heating rate (blue) and the corresponding $\delta$s for one particular field line as a function of time. The second panel shows the zoomed-in coronal integrated viscous heating rate for the time range marked in the top panel. The third panel shows the coronal averaged temperature as a function of time, while the bottom panel shows the energies of the events obtained from Methods B and C in blue and red, respectively. Note that zero energy is slightly offset from the baseline in the bottom plot. Most of the event energies from Method C are very close to zero on this scale. The lower and upper horizontal blue lines are the energy thresholds above which events detected by Method B are considered to be medium and high energy, respectively. Red horizontal lines indicate the same for Method C. Hence, among the three events from Method B that are shown in the zoomed-in time range, two are high-energy nanoflares and the other one is a low energy nanoflare. According to the global delay estimate for events of all energies, all three of these are high frequency nanoflares. 

High frequency nanoflares are typically thought to produce approximately steady temperature. This is the case when the nanoflares have similar energy, but nanoflares in our simulation and on the real Sun exhibit a broad range of energies. The temperature curve in Figure \ref{example_high_energy} is far from steady. It exhibits an evolution expected from a low-frequency nanoflare occuring at $t = 675$ s. The temperature suddenly rises and then slowly declines, except for a small temporary increase from the weak event at $t = 1010$ s. The decline ensues until another large event occurs at $t = 1448$ s.  Nanoflares from Method C also show that, even though the events are high frequency in nature, the overall temperature evolution is determined by the high energy events, which occur at low frequency. Emphasis should also be given on the fact that the temperature evolution not only depends on the amount of energy released but also on the local plasma density. For instance, compared to low density regions, extremely dense regions require more energy to raise the temperature by the same amount.

This brings to the conclusion that, even though nanoflares typically repeat at high frequencies, the temperature evolution in individual magnetic strands is significantly dominated by high energy nanoflares which repeat at low frequencies.

\section{Summary and Discussion}\label{sec:Summary}

This study investigated the frequency and energy distributions of nanoflares that occur self-consistently in a 3D MHD simulation of a subset of an active region \citep{Johnston_2025arXiv250812952J}. Using an automated approach, we tracked 3136 field lines over 6000 s of evolution and characterized the nanoflares that occur along the individual field lines.
We have used three different methods to define a `nanoflare'. 
Each method has its own advantages and disadvantages, which can only be fully understood by carefully studying individual reconnection events in detail.
Our key findings are as follows.

\textit{Nanoflare energy distributions} -- The energy distributions of the nanoflares exhibit a general trend towards a log-normal distribution, differing from the power-law distributions reported previously for full-size flares (\cite{Aschwanden_2025arXiv250318136A} and references therein). Our energies are measured per unit cross-sectional area whereas full-size flare energies are volume integrated. In our simulation, complex footpoint motions and coronal reconnection cause the magnetic flux to fragment, resulting in the formation of current sheets and the reconnection of magnetic field lines. Previous studies indicate that this fragmentation process can lead to the observed log-normal distribution \citep{Bogdan_1988ApJ...327..451B}. 

It is important to note that, for each method, the total energy contained in the nanoflares is less than the total viscous heating that occurs in the simulation. There are several reasons for this. Method A often does not identify the full nanoflare duration. Methods A and B both miss many small events that are captured by Method C, either because the $\delta$ jump of the reconnection is below the adopted threshold or because the field line is heated by the outflow jet from a nearby reconnection site and does not itself participate directly in a reconnection. Some instances of heating of the latter type may be captured by Method B and even Method A if the secondary heating falls with the defined nanoflare duration. As shown in Figure \ref{methodABC}, there seems to be a component of low-level `background' heating that even Method C does not identify. We make no attempt to subtract any background heating when determining nanoflare energies.

Despite this, the `rollover' observed in the lower-energy end of the nanoflare energy distributions from all three methods suggests that any undetected low-energy events contribute minimally to the total energy budget. Instead, the total energy is dominated by intermediate-energy events, which are well captured by the proposed methods.

\textit{Nanoflare occurrence frequencies} -- Our simulation suggests that a broad distribution of nanoflare frequencies heat and maintain the coronal plasma, consistent with observations. Comparison of nanoflare delays with the corresponding cooling timescales is used to characterize the nanoflare frequencies. We find that nanoflares with different energies repeat at different frequencies. Specifically, low energy nanoflares occur at high frequencies and high energy nanoflares at low frequencies. 

Misleading conclusions can be drawn when the energy dependence of the frequency is not taken into account. When the nanoflares are treated in aggregate, without regard to energy, the frequency is high, which is often associated with steady temperatures. However, the temperature evolution of individual field lines is largely driven by high energy nanoflares. Low energy nanoflares produce small deviations to the overall cooling that follows high energy events. 

One possible interpretation of this result is that weak, high frequency nanoflares occur ubiquitously across all field lines at all times to produce the diffuse component of an active region. Superimposed on these high frequency nanoflares, are strong, low frequency nanoflares that occur only on selected field lines when conditions conducive to a nanoflare storm are met \citep{Klimchuk_2015}. These storms give rise to the brighter, more structured loop-like features. Alternatively, these high-energy nanoflares can also occur `randomly' when a field line manages to build up enough magnetic stress, without the need for coordinated storm-like behavior. Definitive distinction between these two scenarios is beyond the scope of the current study and this will be the focus of future investigation.

The presence of a broad range of heating frequencies found in this study is in agreement with the recent work by \cite{Mondal_2025ApJ...980...75M}, where they examined the spatio-temporal evolution of nanoflare heating in an active region by using both observations and hydrodynamic simulations. 
Their results confirm the presence of nanoflares with all frequencies (high, intermediate, and low).
They concluded that while high-frequency nanoflares are the most numerous, they account for less than half of the total energy deposition and primarily contribute to relatively cool plasma ($\sim 1$ MK).

In this paper, we studied the heating on a large number of individual field lines that were identified by a set of random footpoint positions in the photospheric driving plane. The nanoflares occurring on these field lines could be independent events, or they could also be part of a collective phenomenon such as a nanoflare storm that has an avalanche nature. We have suggested that the diffuse component of the observed corona is associated with independent nanoflares, whereas the bright loop component is associated with such nanoflare storms \citep{Klimchuk_2023ApJ...942...10K, Johnston_2025arXiv250812952J}. 
It seems likely that a single field line would experience multiple reconnections during a storm, and we indeed find instances where the $\delta$ exceeded the threshold multiple times in a very short period. 
We plan to investigate the details of nanoflare storms in a future study.

\appendix

\section{Slippage vs. Reconnection in Magnetic Field Line Evolution}
\label{slip_recn}

\begin{figure*}
 \centering
 \includegraphics[width=0.9\textwidth,angle=0]{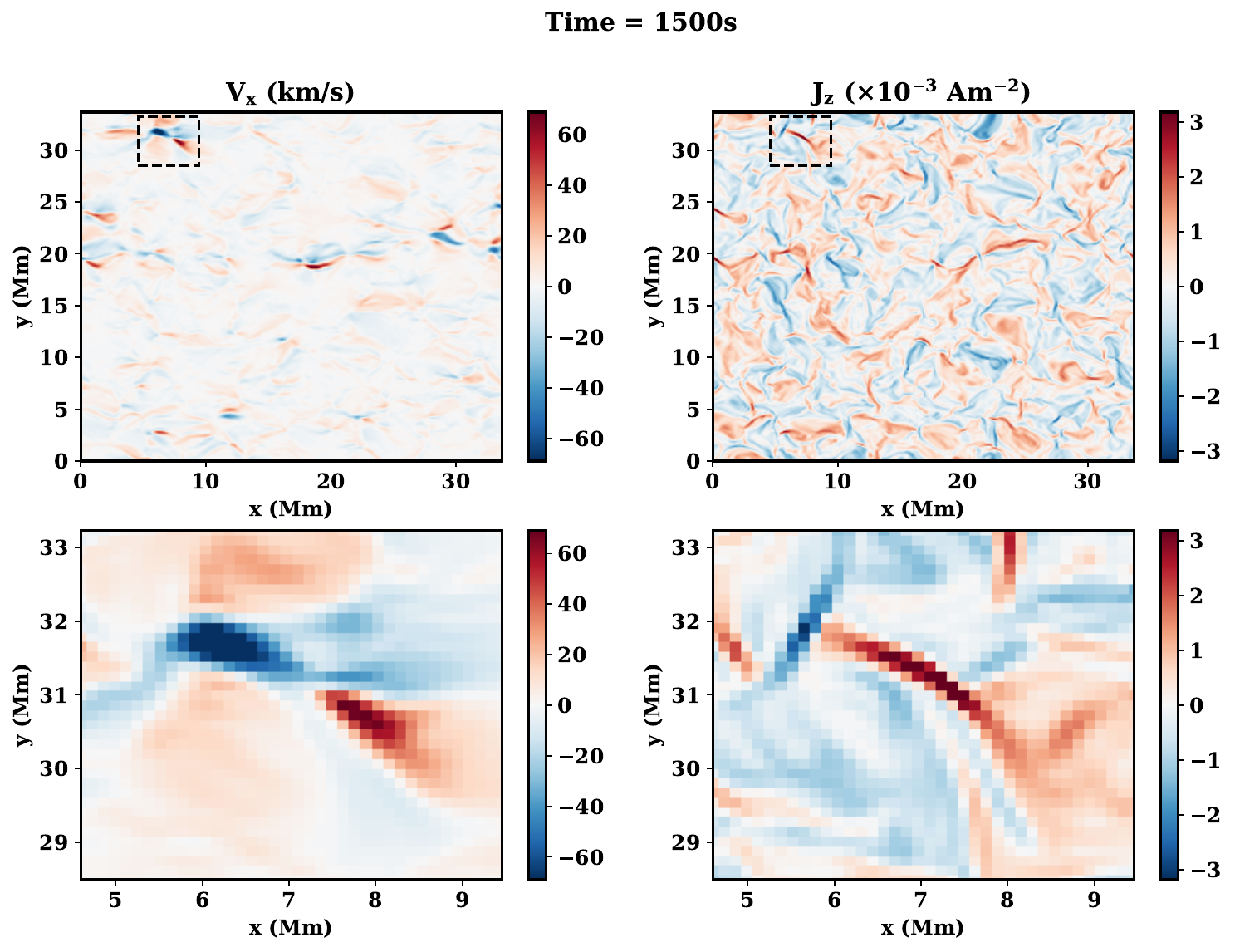}
 \caption{Snapshot of the coronal z-midplane highlighting multiple reconnection sites, characterized by bi-directional flows ($v_x$) and enhanced current density ($J_z$). The bottom panel presents a zoomed-in view of the reconnection region, outlined by the black rectangular box in the top panel. The color maps are saturated to values shown in the corresponding colorbars. Actual peak values are larger.}
 \label{reconnection_site}
 \end{figure*}
 
\begin{figure*}
 \centering
 \includegraphics[width=0.97\textwidth,angle=0]{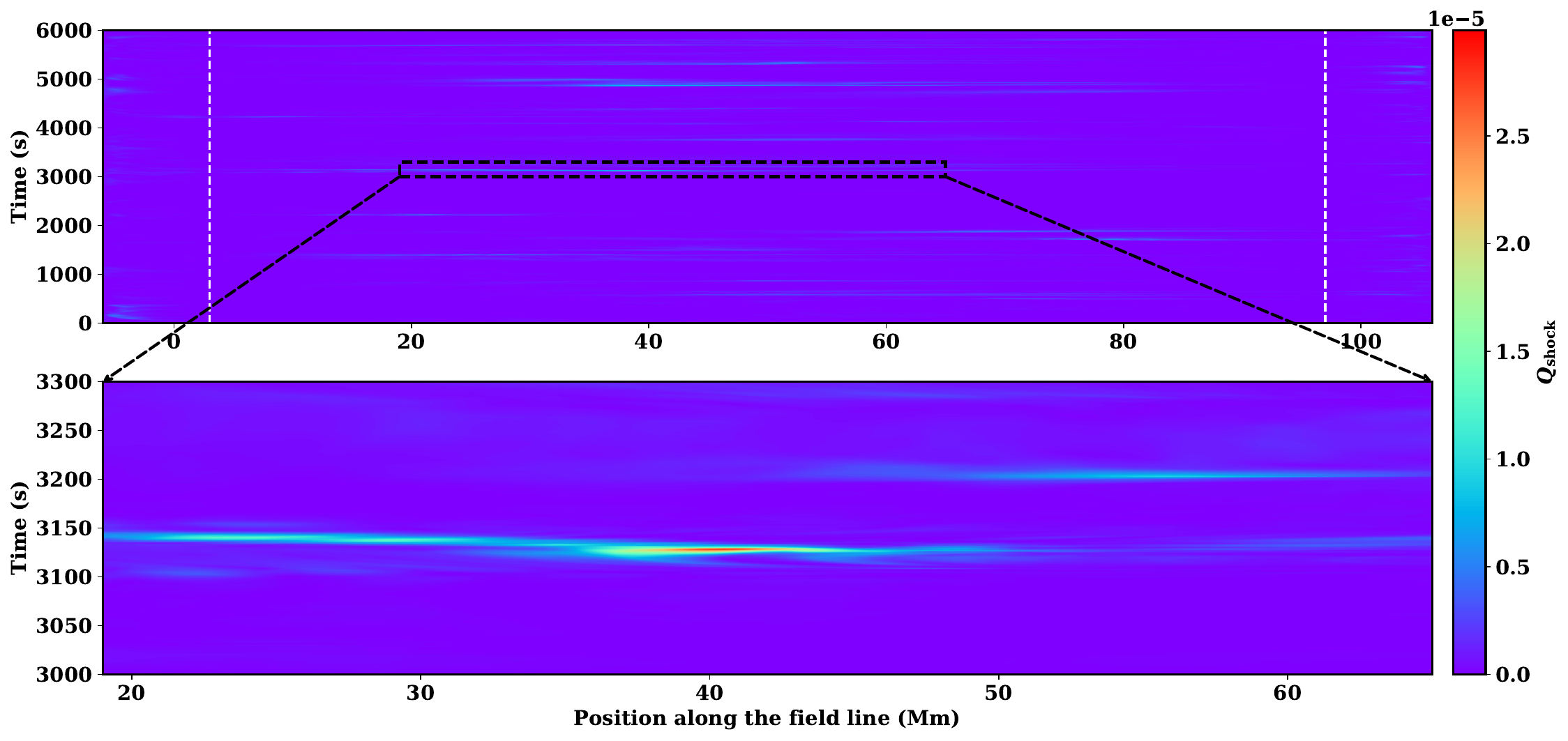}
 \caption{Temporal evolution of the viscous shock heating rate ($Q_{\mathrm{shock}}$) along a randomly chosen magnetic field line is shown in the top panel. The bottom panel provides a zoomed-in view of the localized heating near the field-line apex. The white dashed vertical lines indicate the approximate coronal extent of the selected field line.}
 
 \label{time_evolution_pt1}
 \end{figure*}

While we interpret the measured $\delta$ jumps as resulting from abrupt connectivity changes during magnetic reconnection, it is natural to ask whether they could instead arise from field line slippage due to finite numerical resolution. Disentangling the effects of magnetic field line slippage from reconnection in MHD simulations is inherently challenging, especially in the absence of a prescribed, explicit resistivity. Since the numerical resistivity is both spatially and temporally varying -- likely dependent on local gradients in the magnetic field -- its contribution to field line diffusion is difficult to quantify precisely. Fortunately, there are ways to qualitatively assess the relative importance of true reconnection and numerical slippage in our simulation.

From the magnetic induction equation, the diffusion term that governs field line slippage is proportional to $\nabla^2 \mathbf{B}$. Noting that $\mathbf{J} \propto (\nabla \times \mathbf{B}$) and $\nabla \cdot \mathbf{B} = 0$, it follows that $\nabla^2 \mathbf{B} \propto (\nabla \times \mathbf{J})$. In particular, we consider $(\nabla \times \mathbf{J})_{\perp}$, evaluated with respect to the local magnetic field direction, as a measure of field line slippage, since magnetic diffusion acts transverse to the field lines. If $\delta$ is dominated by slippage, then it should be closely correlated with $(\nabla \times \mathbf{J})_{\perp}$. The correlation should be one-to-one, such that $\delta$ is large when $(\nabla \times \mathbf{J})_{\perp}$ is large, and small when $(\nabla \times \mathbf{J})_{\perp}$ is small. 

The same one-to-one correspondence does not hold for magnetic reconnection. While reconnection occurs at current sheets--regions where $(\nabla \times \mathbf{J})_{\perp}$ is large--not all current sheets are actively reconnecting at a given time. Therefore, although a large reconnection $\delta$ typically coincides with strong $(\nabla \times \mathbf{J})_{\perp}$, the reverse is not always true: $\delta$ can be small even when $(\nabla \times \mathbf{J})_{\perp}$ is large. This distinction arises because most current sheets evolve gradually in a quasi-equilibrium state until certain onset conditions for reconnection are met. The physics of onset is actively being studied, but it could involve a critical thickness \citep{Leake_2020ApJ...891...62L, Leake_2024ApJ...973...21L} or a critical shear \citep{Klimchuk_2023FrP....1198194K}. Only once these conditions are satisfied do two field lines reconnect, producing a non-zero $\delta$. In contrast, slippage driven by numerical diffusion results in a large $\delta$ whenever $(\nabla \times \mathbf{J})_{\perp}$ is large. This difference in behavior provides a clear discrimination between reconnection-driven and slippage-driven $\delta$s. 

There are additional clear signatures that distinguish reconnection from slippage. Magnetic reconnection generates fast, bidirectional outflow jets as newly reconnected field lines snap away from the reconnection site. In contrast, any flows associated with slippage are much slower, as field lines `slip' through the plasma rather than exerting significant forces on it. Fast reconnection outflows produce steep velocity gradients, leading to strong localized viscous heating. So we expect strong viscous heating with reconnection but not  with slippage. 

We see all the signatures of reconnection in our simulation. Figure \ref{reconnection_site} shows a snapshot after the system achieves statistical steady state. The top panel displays color maps of the horizontal velocity component ($V_{x}$) and vertical current density ($J_z$) in the coronal mid-plane at $Time = 1500$ s. The bottom panel presents a zoomed-in view of the region enclosed by black boxes in the top panel, highlighting a localized area of interest. Within this region, we observe clear signatures indicative of magnetic reconnection. In particular, the $V_x$ maps reveal bi-directional plasma flows, consistent with reconnection outflow jets. This flow structure aligns spatially with a region of intensified $J_z$, consistent with the presence of a thin current sheet.

Figure \ref{time_evolution_pt1} shows the time evolution of viscous shock heating rate ($Q_{\mathrm{shock}}$) for a randomly selected field line traced from the lower footpoint. The $y$-axis denotes time, while the $x$-axis corresponds to the position along the field line. The white dashed vertical lines demarcate the approximate coronal portion of the field line. As evident, the heating events are short-lived and are concentrated primarily in the coronal portion of the field line. The bottom panel shows a zoomed-in view of a heating event concentrated at the field line apex.

\begin{figure*}
 \centering
 \includegraphics[width=0.95\textwidth,angle=0]{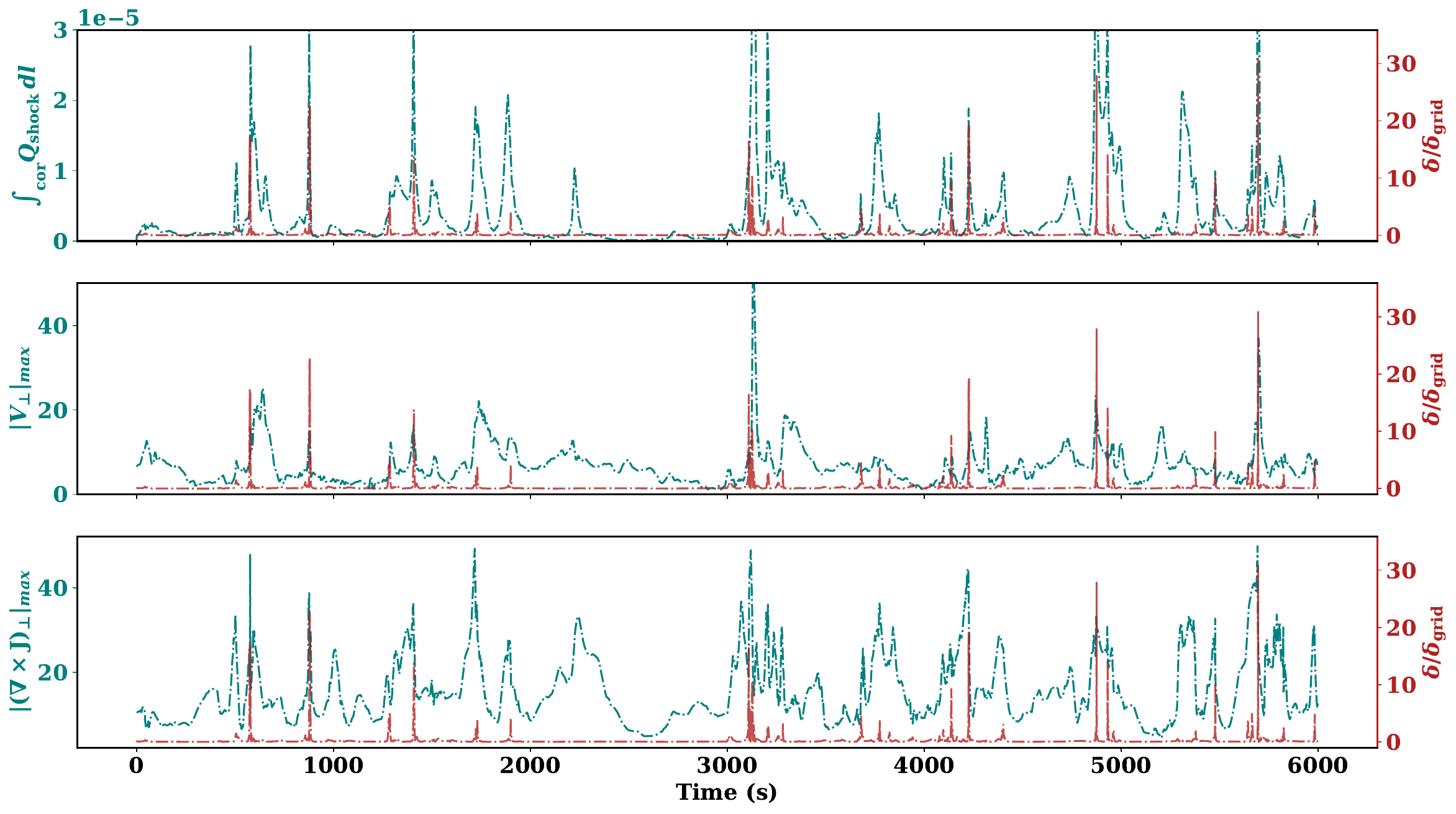}
 \caption{The teal curves show various plasma parameters evaluated along a single magnetic field line. The top panel displays the time evolution of the coronal line-integrated shock heating, $Q_{\mathrm{shock}}$, for the field line shown in Figure~\ref{time_evolution_pt1}, corresponding to the integration range marked by the white dashed lines in that figure. The middle panel shows the temporal variation of the maximum perpendicular velocity component, $v_{\perp}$, within the coronal segment of the same field line. Similarly, the bottom panel presents the evolution of the maximum value of the perpendicular component of the curl of current density, $(\nabla \times \mathbf{J})_{\perp}$, along the coronal portion of the same field line. In each panel, the red curve indicates the corresponding normalized reconnection jump, labeled to the right y-axis. }
 \label{line_plots_pt1}
 \end{figure*}

Figure \ref{line_plots_pt1} shows the time evolution of four quantities for one particular field line tracked from the lower boundary. The teal curves in the panels are, from top to bottom, the coronal-integrated $Q_{\mathrm{shock}}$, the maximum of $V_{\perp}$ along the coronal portion of field line, and the maximum of $(\nabla \times \mathbf{J})_{\perp}$ along coronal portion. The red curve is $\delta$ and is the same in all panels. We see that $\int_{cor}Q_{\mathrm{shock}} dl$, $V_{\perp,max}$, and $(\nabla \times \mathbf{J})_{\perp,max}$ are all large when $\delta$ is large. There also intervals, some very extended, when $(\nabla \times \mathbf{J})_{\perp,max}$ is large but $\delta$ is small -- in contradiction with the diffusion scenario -- such as $Time = 2000 - 3000$ s. This interval includes an event of strong heating and velocity without a large $\delta$ ($Time = 2250$ s), likely due to a nearby reconnection. If the estimated $\delta$ were due to slippage, we would not expect to see such peaks in the intervals where $\delta$ remains very small. Hence, the events in our simulation strongly favor the reconnection interpretation. Lastly, we remind the reader that the small asymmetry in the plot of $\delta$ versus the height of maximum $Q_{\mathrm{shock}}$ (Figure \ref{delta_vs_ht}) is yet another indicator that reconnection dominates over slippage, as discussed in Section \ref{sec:coronal_recn}. 

To summarize, the discriminators between true magnetic reconnection and numerical slippage of field lines are as follows. With reconnection, large $\delta$ should be accompanied by large $(\nabla \times \mathbf{J})_{\perp}$, large $V_{\perp}$, and large $Q_{\mathrm{shock}}$. With slippage, large $\delta$ should be accompanied by large $(\nabla \times \mathbf{J})_{\perp}$, but not necessarily large $V_{\perp}$ or large $Q_{\mathrm{shock}}$. The strongest evidence that reconnection dominates is the presence of small $\delta$ when $(\nabla \times \mathbf{J})_{\perp}$ is large. This would not occur with slippage. Finally, we note that $V_{\perp}$ and $Q_{\mathrm{shock}}$ can sometimes be large even when $\delta$ is small in a reconnection dominated system. Field lines that do not reconnect can be affected by outflow jets from nearby reconnection sites, thereby acquiring $V_{\perp}$ and $Q_{\mathrm{shock}}$ without a $\delta$.

\section{Relative Contribution of Nanoflares To Total Heating}
\label{energy_fraction}
\begin{figure*}
 \centering
 \includegraphics[width=0.9\textwidth,angle=0]{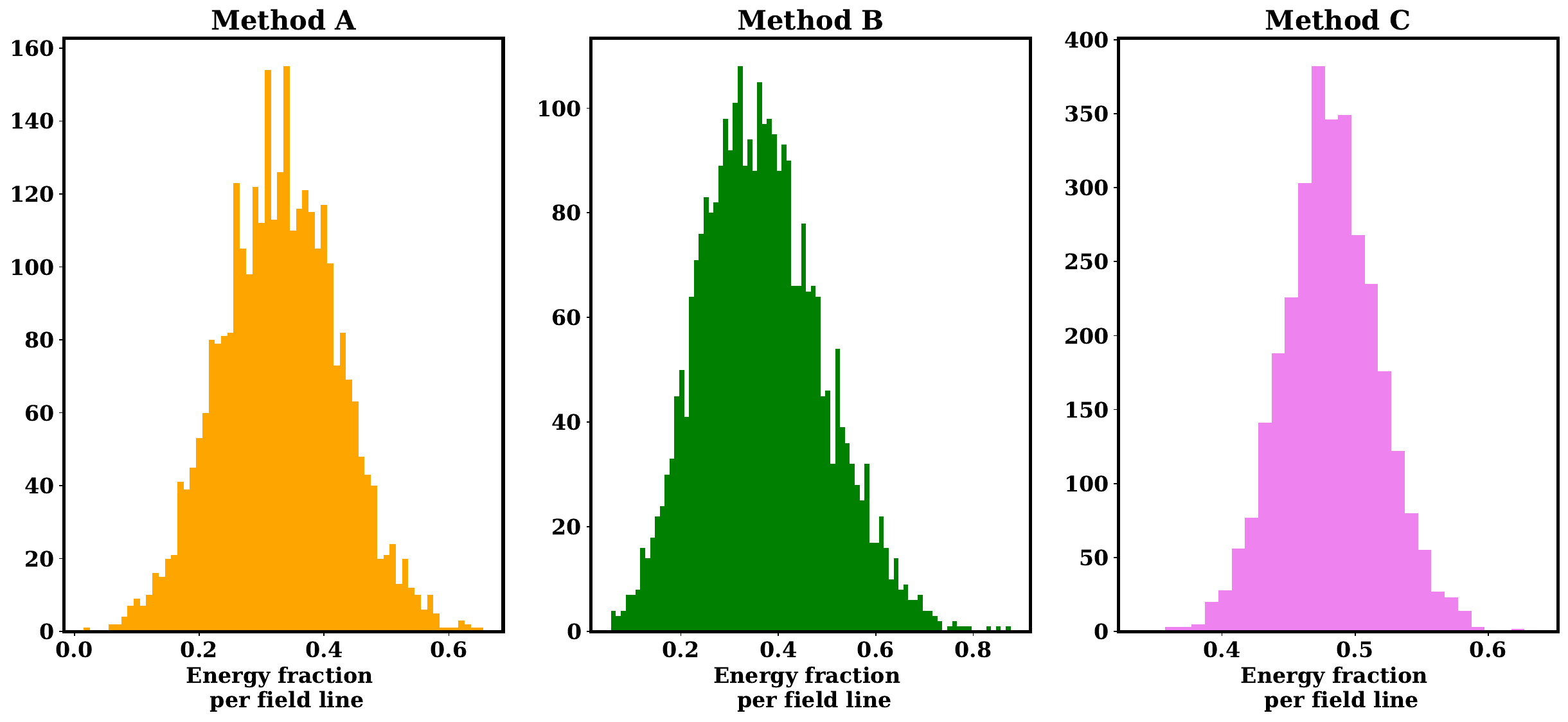}
 \caption{Histograms illustrating the relative fraction of nanoflare energy compared to the total energy from viscous shock heating per field line. The histograms indicate that nanoflares identified by Method A contribute at least 30\% of the total energy per field line, whereas those from Methods B and C contribute at least 40\% and 50\%, respectively.}
 \label{fig:energy_fraction}
 \end{figure*}

\begin{figure}
 \centering
 \includegraphics[width=0.5\textwidth]{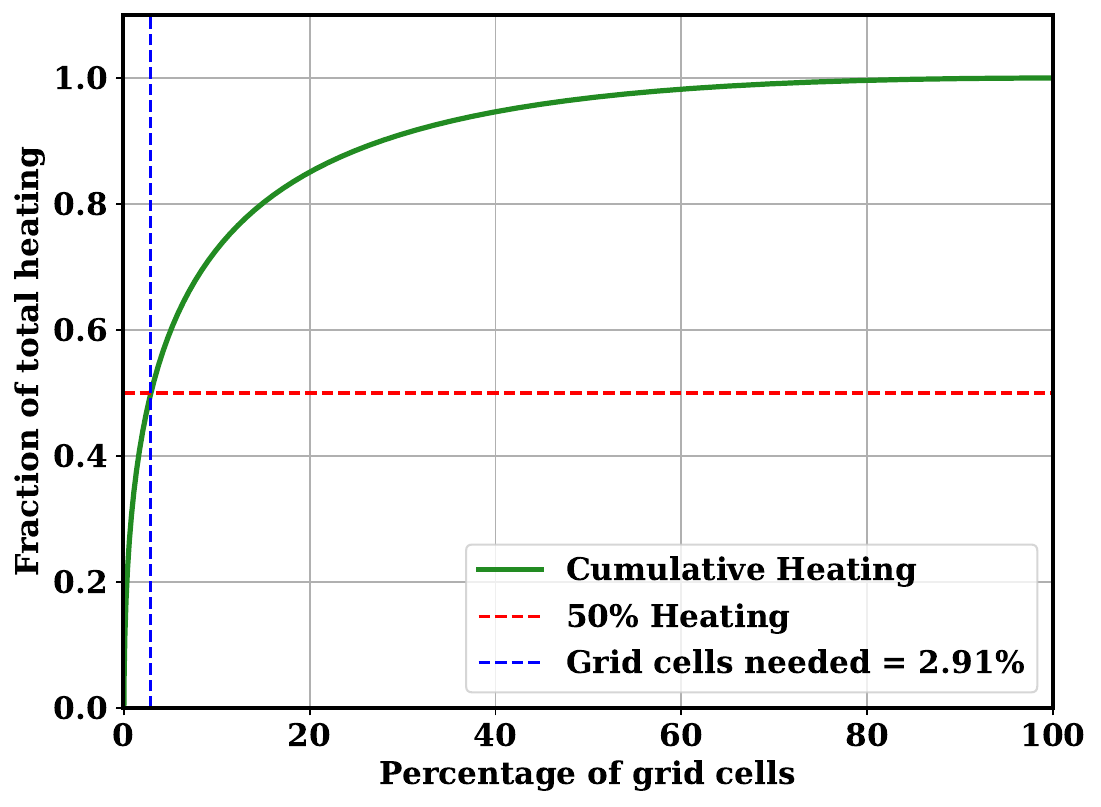}
  \caption{Cumulative viscous shock heating versus percentage of contributing grid cells. At this instant ($t = 1500$ s), less than 3\% of the grid cells account for 50\% of the total heating. This behavior is consistent at all times once the system achieves statistical steady state.}
 \label{grid_fraction}
 \end{figure}

For each field line, we estimate the relative fraction of nanoflare heating compared to the total viscous shock heating along the field line for the entire simulation time. Figure \ref{fig:energy_fraction} shows the histograms of nanoflare heating fractions per field line as obtained from Methods A, B and C. As evident from the figure, the estimated relative contributions of nanoflare heating to the total are roughly 30\% (Method A), 40\% (Method B), and 50\% (Method C). These values represent lower limits of the nanoflare energy content. For instance, Method A clearly underestimates the nanoflare duration, as we have discussed. Even Method B, which defines event durations using the FWHM of the peak heating rate, excludes the extended tails of the heating profile, which can constitute a significant fraction of a nanoflare’s total energy. An important point is that most of the heating is localized. Only a small fraction of the domain contributes most significantly to the total heating: approximately 3\% of the grid cells account for 50\% of the total viscous heating. This is illustrated in Figure \ref{grid_fraction}, which shows the cumulative viscous shock heating as a function of the fraction of contributing grid cells at $t = 1500$ s. We estimated the cumulative heating by ranking the grid cells from strongest to weakest in heating rate and progressively summing their contributions. As the figure demonstrates, at any given time, the top 25\% of grid cells contribute nearly 90\% of the total heating on average. Furthermore, as shown in Figure \ref{example_high_energy}, the plasma thermal evolution is governed mostly by the largest events which are well captured by all these three methods.

 \begin{figure*}
 \centering
 \includegraphics[width=0.98\textwidth,angle=0]{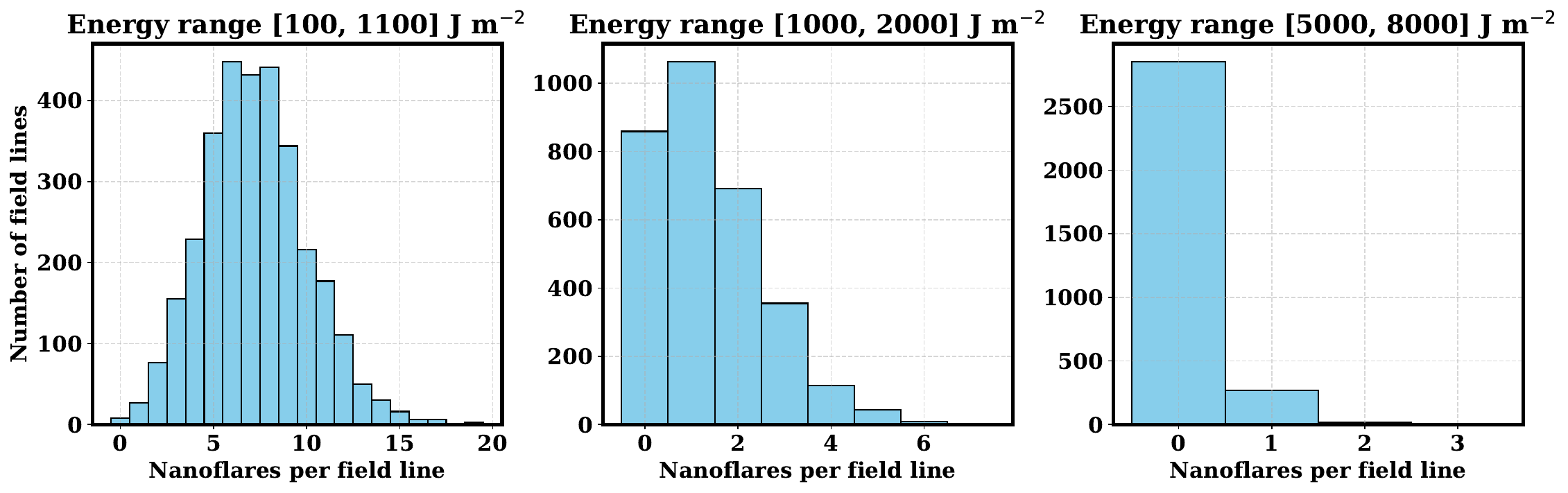}
 \caption{Histograms showing the number of nanoflares occurring per field in different energy ranges. The nanoflares in this example are defined using Method B.}
 \label{fig:nanoflares_per_fl}
 \end{figure*}

\section{Number of Nanoflares per field line}
\label{nanoflares_per_fl}
From the nanoflare energy distributions (Figure \ref{three_methods_fitting}), we inferred that low energy nanoflares repeat at higher frequencies compared to the high energy nanoflares. Although this conclusion was drawn from the entire domain, the results are expressed on a per–field-line basis. The actual frequency on individual field lines will be comparable to the global average if all field lines have similar heating properties. We estimated the number of nanoflares per field line in different ranges which supports this interpretation. Figure \ref{fig:nanoflares_per_fl} shows the histograms of nanoflare counts per field line -- for three different energy ranges -- obtained from Method B. The low energy nanoflares (100–1100 J m$^{-2}$; left panel) are relatively common across many field lines, with counts per field line peaking around 6–8 events. Medium-energy nanoflares (1000–2000 J m$^{-2}$; middle panel) are less frequent, with most field lines hosting one or two such events. High-energy nanoflares (5000–8000 J m$^{-2}$; right panel) are further less frequent, occurring in only a small fraction of field lines. This supports our global estimate that high energy nanoflares occur less frequently, whereas low and medium energy nanoflares occur at higher and intermediate frequencies, respectively. 

\section*{Acknowledgements}
SSM, LKSD, JAK, and CDJ were funded by NASA’s Internal Scientist Funding Model (competed work package program) at Goddard Space Flight Center. Simulations were performed on NASA’s High End Computing Facilities by CDJ. Source code, raw data, and plotting routines are available upon request from the authors. We would like to thank James E. Leake, N. Dylan Kee and Yi-Min Huang for fruitful discussions. 
We sincerely thank the anonymous referee for their constructive comments and valuable suggestions, which have helped improve the quality and clarity of this manuscript.

\bibliography{ms}

\end{document}